\documentclass[11pt,oneside,reqno]{amsart}
\usepackage[foot]{amsaddr}
\usepackage[margin=1in]{geometry}
\usepackage{natbib}
\allowdisplaybreaks

\RequirePackage[OT1]{fontenc}
\RequirePackage{amsthm,amsmath,amssymb}
\RequirePackage{graphicx,bbm,booktabs,xr}
\RequirePackage{natbib}
\RequirePackage[colorlinks,citecolor=blue,urlcolor=blue]{hyperref}

\newcommand{\R}{\mathbb{R}}
\newcommand{\E}{\mathbb{E}}
\renewcommand{\Pr}{\operatorname{Pr}}
\newcommand{\argmax}{\operatorname{argmax}}
\newcommand{\argmin}{\operatorname{argmin}}

\newcommand{\N}{\mathcal{N}}

\newcommand{\true}{\text{true}}

\newcommand{\defeq}{:=}

\newcommand{\appref}[1]{Appendix~{\ref{sec:#1}}}
\newcommand{\appsref}[1]{Appendices~{\ref{sec:#1}}}
\newcommand{\appssref}[1]{{\ref{sec:#1}}}

\newcommand{\eqnref}[1]{Eq.~\ref{eqn:#1}}
\newcommand{\eqnsref}[1]{Eqs.~\ref{eqn:#1}}
\newcommand{\eqnssref}[1]{\ref{eqn:#1}}

\newcommand{\secref}[1]{Section~{\ref{sec:#1}}}

\begin{document}

\title[Bayesian Analysis of Simultaneous Changepoints]
{Empirical Bayesian Analysis of Simultaneous Changepoints in Multiple
Data Sequences}
\author{Zhou Fan}
\author{Lester Mackey}
\address{Department of Statistics, Stanford University and Microsoft Research}
\email{zhoufan@stanford.edu, lmackey@microsoft.com}
\thanks{ZF was supported by a Hertz Foundation Fellowship and an NDSEG
Fellowship (DoD, Air Force Office of Scientific Research, 32 CFR
168a). LM was supported by a Terman Fellowship.}

\begin{abstract}
Copy number variations in cancer cells and volatility fluctuations in stock
prices are commonly manifested as changepoints occurring at the same positions
across related data sequences. We introduce a Bayesian modeling framework,
BASIC, that employs a changepoint prior to capture the co-occurrence tendency
in data of this type.
We design efficient algorithms to sample from and maximize over the BASIC changepoint posterior
and develop a Monte Carlo expectation-maximization procedure
to select prior hyperparameters in an empirical Bayes fashion.
We use the resulting BASIC framework to analyze
DNA copy number variations in the NCI-60 cancer cell lines and to identify
important events that affected the price volatility of S\&P 500 stocks from
2000 to 2009.
\end{abstract}
\maketitle

\section{Introduction}\label{sec:introduction}
\begin{figure}[tbp]
\includegraphics[width=\textwidth]{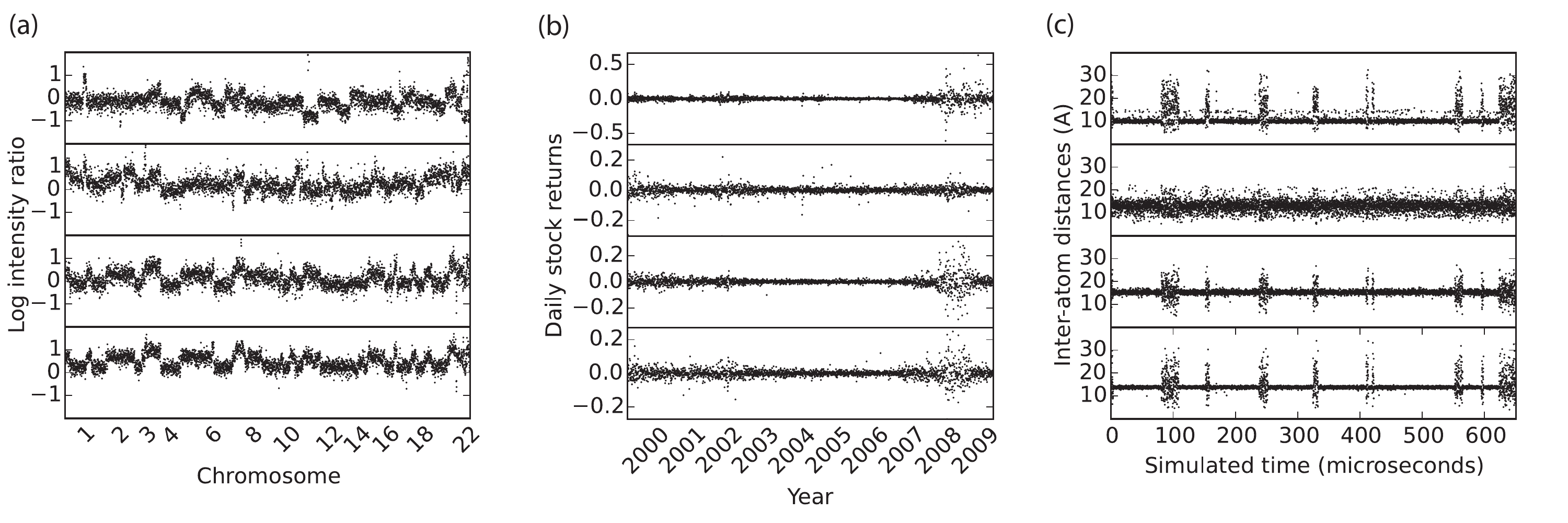}
\caption{(a) DNA copy numbers in four cancer cell lines,
indicated by fluorescence intensity log-ratios
from array-CGH experiments. 
(b) Daily returns of four U.S. stocks.
(c) Distances between four pairs of atoms in a computer
simulation of a protein molecule.}\label{figrealexamples}
\end{figure}

Figure \ref{figrealexamples} displays three examples of aligned sequence
data. Panel (a) presents DNA copy number measurements at sorted genome locations
in four human cancer cell lines \citep{varmaetal}. Panel (b)
shows the daily stock returns of four U.S. stocks over a period of ten years. 
Panel (c) traces the interatomic distances between four pairs of atoms in a protein 
molecule over the course of a computer simulation \citep{lindorfflarsenetal}.
Each sequence in each panel is reasonably modeled as having a number of
discrete ``changepoints,'' such that the characteristics of the data
change abruptly at each changepoint but remain homogeneous between changepoints.
In panel (a), these changepoints demarcate the
boundaries of DNA stretches with abnormal copy number. In panel (b),
changepoints indicate historical events
that abruptly impacted the volatility of stock returns. In panel (c),
changepoints indicate structural changes in the 3-D conformation of the protein 
molecule. For each of these examples, it is important to understand
when and in which sequences changepoints occur. 
However, the number and locations of these changepoints are typically not known 
a priori and must be estimated from the data. The problem of detecting
changepoints in sequential data has a rich history in the statistics literature,
and we refer the reader to \citep{bassevillenikiforov,chengupta} for a more
detailed review and further applications.

In many modern applications, we have available not just a single data sequence
but rather many related sequences measured at the same locations or time points. These sequences often exhibit changepoints occurring at the same sequential locations.
For instance, copy number variations frequently occur at common genomic 
locations in cancer samples \citep{pollacketal} and in biologically-related
individuals \citep{zhangetal},
economic and political events can impact the volatility of 
many stock returns in tandem, and a conformational change in a region
of a protein molecule can affect
distances between multiple atomic pairs \citep{fanetal}. As recognized in
many recent papers, discussed below, an analysis of multiple sequences
jointly may yield greater statistical power in detecting their changepoints
than analyses of
the sequences individually. In addition, a joint analysis may
more precisely identify the times or locations at which changepoints occur
and better highlight the locations where changepoints most frequently recur
across sequences.

Motivated by these considerations,
we introduce a Bayesian modeling framework, \textbf{BASIC}, for carrying out a
\textbf{B}ayesian \textbf{A}nalysis of \textbf{SI}multaneous
\textbf{C}hangepoints.
In single-sequence applications, Bayesian changepoint detectors have been shown to exhibit favorable
performance in comparison with other available methods and have enjoyed
widespread use \citep{chernoffzacks,yao,barryhartigan,stephens,chib,fearnhead,
adamsmackay}.
In Section~\ref{sec:model}, we propose an extension of Bayesian changepoint detection to the
multi-sequence setting by
defining a hierarchical prior over latent changepoints, which first
specifies the sequential locations at which changepoints may occur
and then specifies the sequences that contain a changepoint at each such
location.

Inference in the BASIC model is carried out through efficient, tailored Markov
chain Monte Carlo (MCMC) procedures (Section \ref{subsecsampling})
and optimization procedures (Section \ref{subsecmaximization}) designed to
estimate the posterior probabilities of changepoint events
and the maximum-a-posteriori (MAP) changepoint locations, respectively.
These procedures employ dynamic programming sub-routines to avoid becoming
trapped in local maxima of the posterior distribution.
To free the user from pre-specifying prior hyperparameters, we adopt an empirical Bayes approach \citep{robbins1956} to automatic
hyperparameter selection using Monte Carlo expectation maximization (MCEM) \citep{weitanner} (Section \ref{subsecMCEM}).

To demonstrate the applicability of our model across different application
domains, we use our methods to analyze two different data sets. The
first is a set of array comparative genomic hybridization 
(aCGH) copy number measurements of the
NCI-60 cancer cell lines \citep{varmaetal}, four of which
have been displayed in Figure \ref{figrealexamples}(a). 
In Section \ref{secCNV}, we
use our method to highlight focal copy number variations that are present in
multiple cell lines; many of the most prominent variations that we detect
are consistent with known or suspected oncogenes and tumor suppressor genes.
The second data set consists of the daily returns of 401 U.S.
stocks in the S\&P 500 index from the year 2000 to 2009, four of which have been
displayed in Figure \ref{figrealexamples}(b). In Section \ref{secstocks}, we use our method to
identify important events in the history of the U.S.\
stock market over this time period, pertaining to the entire market as well as
to individual groups of stocks.

{\bf Comparison with existing methods: } Early work on changepoint detection
for multivariate data \citep{srivastavaworsley,healy} studied the detection
of a change in the joint distribution of all observed variables. Our viewpoint
is instead largely shaped by \citep{zhangetal}, which formulated the problem as
detecting changes in the marginal distributions of subsets of these variables.
A variety of methods have been proposed to address
variants of this problem, many with a particular focus on analysis of
DNA copy number variation. These methods include segmentation
procedures using scan statistics \citep{zhangetal,siegmundetal,jengetal},
model-selection penalties \citep{zhangsiegmund,fanetal},
total-variation denoising \citep{gnowaketal,zhouetal}, and
other Bayesian models \citep{dobigeonetal,shahetal,harleetal,
bardwellfearnhead}. Here, we briefly highlight several
advantages of our present approach.

Comparing modeling assumptions, several methods
\citep{jengetal,bardwellfearnhead} focus on the
setting in which each sequence exhibits a baseline behavior,
and changepoints demarcate the boundaries of non-overlapping
``aberrant regions'' that deviate from this baseline. 
\cite{shahetal} further assumes a hidden Markov model with a small finite set
of possible signal values for each sequence.
However, data in many applications are not well-described by these simpler
models. For instance, in cancer samples, short focal copy number aberrations
may fall inside
longer aberrations of entire chromosome arms and overlap in
sequential position, and true copy numbers
might not belong to a small set of possible values if there are fractional
gains and losses due to sample heterogeneity. Conversely, the Bayesian models of
\citep{dobigeonetal,harleetal} are very general, but their priors
and inference procedures involve $2^J$ parameters (where $J$
is the number of sequences), rendering inference intractable for
applications with many sequences. By introducing a
prior that is exchangeable across sequences, we strike a
different balance between model generality and tractability of inference.

Comparing algorithmic approaches, we observe in simulation (Section 
\ref{sec:simulation}) that total-variation denoising can severely
overestimate the number of changepoints, rendering them ill-suited for
applications in which changepoint-detection accuracy (rather than 
signal reconstruction error) is of interest. In contrast to recursive
segmentation procedures, our algorithms employ sequence-wise local moves, which
we believe are better-suited to multi-sequence problems with
complex changepoint patterns. These local moves are akin to the
penalized likelihood procedure of \citep{fanetal}, but in contrast
to \citep{fanetal} where the likelihood penalty shape and magnitude are
ad hoc and user-specified,
our empirical Bayes approach selects prior hyperparameters automatically using
MCEM. Finally, the BASIC approach provides a unified framework that
accommodates a broad
range of data types and likelihood models, can detect changes of various types
(e.g.\ in variance as well as in mean), and returns posterior probabilities for
changepoint events in addition to point estimates.

\section{The BASIC Model}\label{sec:model}
\begin{figure}[tbp]
\centering
\includegraphics[width=0.9\textwidth]{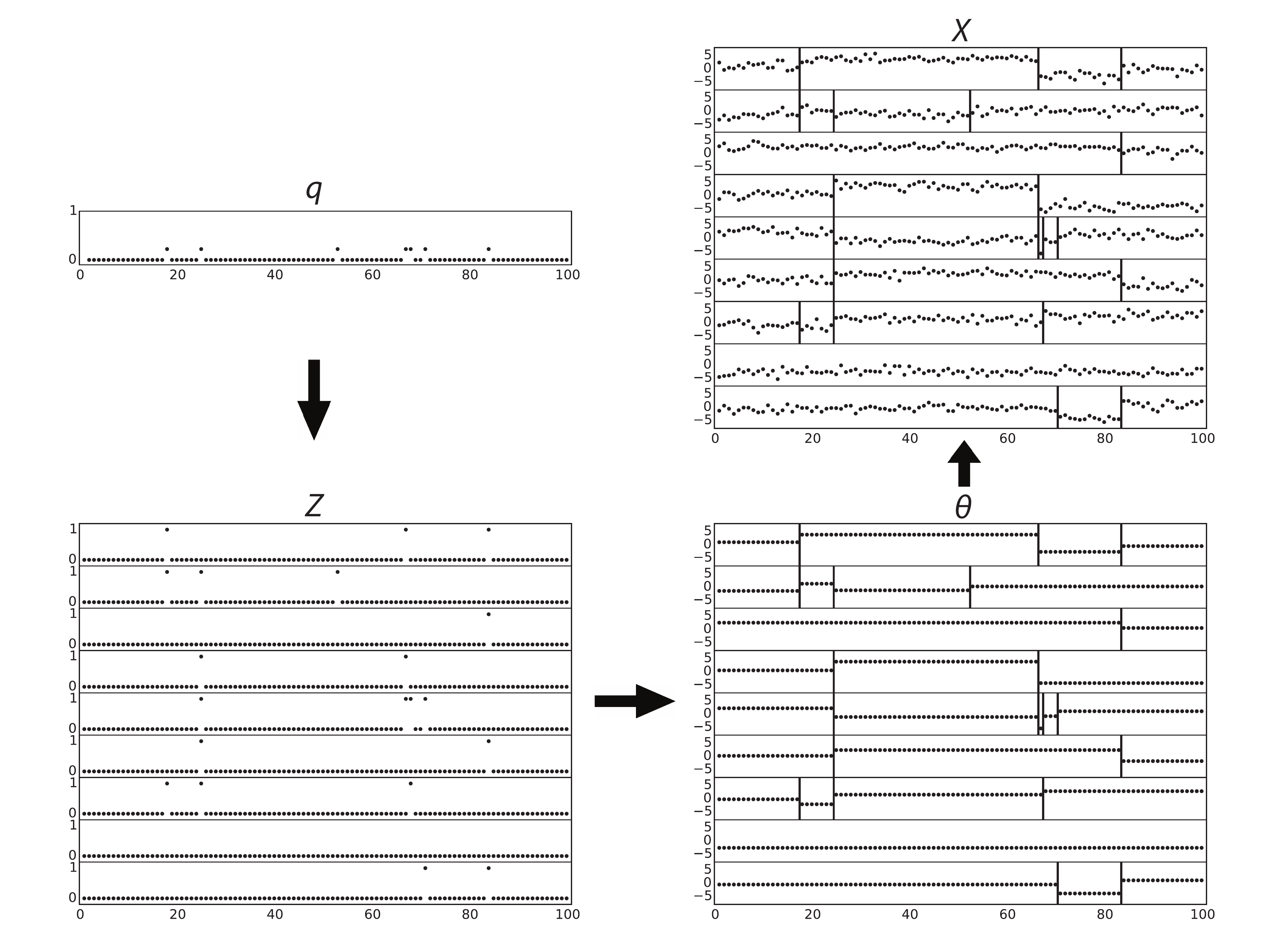}
\caption{An illustration of the BASIC model. 
In this illustration, distinct values of
$\theta$ are drawn from $\pi_\Theta=\operatorname{Normal}(0,5)$, and values of
$X$ are drawn from
$p(\cdot|\theta)=\operatorname{Normal}(\theta,1)$.
}\label{figmodel}
\end{figure}
Suppose $X\in \R^{J \times T}$ is a collection of $J$ aligned data sequences, each consisting of $T$ observations.  
The BASIC model for $X$ is a generative process defined by three inputs: an observation likelihood $p(\cdot|\theta)$ parameterized by $\theta\in\Theta \subseteq \R^d$, a prior distribution $\pi_\Theta$ on the parameter space $\Theta$, and a changepoint frequency prior $\pi_Q$ on $[0,1]$.
For each sequence position $t$, a latent variable $q_t \in [0,1]$ is
drawn from $\pi_Q$ and represents the probability of any
sequence having a changepoint between its $(t-1)^\text{th}$ and
$t^\text{th}$ data points. Then, for each sequence position $t$ and sequence
$j$, a latent variable $Z_{j,t} \in \{0,1\}$ is drawn with
$\Pr[Z_{j,t}=1]=q_t$ and indicates whether there is a changepoint in
sequence $j$ between its $(t-1)^\text{th}$ and $t^\text{th}$ 
data points. Finally, for each
$t$ and $j$, a latent likelihood parameter
$\theta_{j,t} \in \Theta$ and an observed data point $X_{j,t}$ are drawn, such
that $\theta_{j,t}$ remains constant (as a function of $t$) in each data
sequence between each pair of consecutive changepoints of that sequence and
is generated anew from the prior $\pi_\Theta$ at each changepoint, and
$X_{j,t}$ is a conditionally independent draw from $p(\cdot|\theta_{j,t})$.
This process is summarized as follows:\\\\

\begin{center}
{\bf The BASIC Model}
\end{center}
\vspace{-0.1in}
\begin{align*}
q_2,\ldots,q_T &\overset{iid}{\sim} \pi_Q\\
Z_{j,t}|q_t &\overset{ind}{\sim}
\operatorname{Bernoulli}(q_t) &\forall\,j=1,\ldots,J \text{ and } t=2,\ldots,T\\
\theta_{1,1},\ldots,\theta_{J,1} &\overset{iid}{\sim} \pi_\Theta\\
\theta_{j,t}|Z_{j,t},\theta_{j,t-1} &\begin{cases}
\overset{ind}{\sim} \pi_\Theta & \text{if } Z_{j,t}=1\\
=\theta_{j,t-1} & \text{if } Z_{j,t}=0
\end{cases} & \forall\,j=1,\ldots,J \text{ and } t=2,\ldots,T\\
X_{j,t}|\theta_{j,t} &\overset{ind}{\sim} p(\cdot|\theta_{j,t})
& \forall\,j=1,\ldots,J \text{ and } t=1,\ldots,T
\end{align*}
\noindent
For notational convenience, we arrange
$Z_{j,t}$ into a matrix $Z \in \{0,1\}^{J \times T}$, 
fixing $Z_{j,1}=0$ for all $j=1,\ldots,J$. 
Figure \ref{figmodel} illustrates this generative model in the case where the
piecewise-constant parameter $\theta_{j,t}$ represents the mean of the
distribution of $X_{j,t}$, and $X_{j,t}$ is normally-distributed around
this mean with fixed unit variance. Our primary goal in this
model will be to infer the latent changepoint variables $Z$ upon
observing the data $X$.

A key input to the model is the prior distribution $\pi_Q$ over $[0,1]$,
which controls how frequently changepoints occur and to what extent they
co-occur across sequences. Rather than requiring the user to pre-specify
this prior, Section \ref{subsecMCEM} develops an empirical Bayes MCEM procedure to select
$\pi_Q$ automatically. Specifically, we parametrize $\pi_Q$ as a mixture
distribution
\begin{equation}\label{eqn:piq}
\pi_Q=\sum_{k \in S} w_k \nu_k,
\end{equation}
where $\{\nu_k\}_{k \in S}$ is a fixed finite dictionary of probability
distributions over $[0,1]$ and $\{w_k\}_{k \in S}$ are non-negative mixture
weights summing to 1, and the MCEM maximum marginal likelihood procedure selects the weights
$\{w_k\}_{k \in S}$. In our applications, we will simply take the dictionary
$\{\nu_k\}_{k \in S}$ to be discrete point masses
over a fine grid of points in $[0,1]$.

The choices of the likelihood model $p(\cdot|\theta)$ and the prior distribution
$\pi_\Theta$ are application-dependent.
For our analysis of DNA copy number variations in Section \ref{secCNV},
we use a normal model for $p(\cdot|\theta)$ where $\theta$ parametrizes
the normal mean, and $\pi_\Theta$ is the normal conjugate prior.
For our analysis of stock return volatility in Section \ref{secstocks}, we use a
Laplace model for $p(\cdot|\theta)$ with mean 0 and scale parameter $\theta$,
and $\pi_\Theta$ is the inverse-Gamma conjugate prior. We provide details on
these and several other common models in \appref{likelihoodmodels}.
Our inference procedures are tractable whenever the marginal
 \begin{equation}\label{eqn:P}
P_j(t,s):=\int \prod_{r=t}^{s-1} p(X_{j,r}|\theta) \pi_\Theta(d\theta)
 \end{equation}
may be computed quickly from $P_j(t,s-1)$ and $P_j(t-1,s)$. This holds, in
particular, whenever $p(\cdot|\theta)$ is an exponential family model
with $\pi_\Theta$ the conjugate prior, as $P_j(t,s)$ may be computed
by updating a fixed number of sufficient statistics. Any unspecified
hyperparameters of $\pi_\Theta$ can also be selected automatically
using the MCEM procedure of Section \ref{subsecMCEM}.

We have assumed for notational convenience
that each data sequence is generated from the same parametric family
$p(\cdot|\theta)$ with the same prior $\pi_\Theta$. In applications where
sequences represent different types of quantities, the choices of
$p(\cdot|\theta)$ and $\pi_\Theta$ should vary across sequences, and
our posterior inference algorithms are easily extended to accommodate this
setting.

\section{Inference procedures}\label{sec:inference}
In this section, we give a high-level overview of our algorithms for posterior
inference in the BASIC model, deferring details to
\appsref{samplingalgorithms}-\appssref{MCEMalgorithms}.
Our primary task is to perform posterior
inference of the unobserved latent changepoint variables $Z$, given the observed
data $X$. Assuming $\pi_Q$ and $\pi_\Theta$ are fixed and known,
Section \ref{subsecsampling} presents an MCMC procedure for sampling from the
posterior distribution $\Pr(Z|X)$, and
Section \ref{subsecmaximization} presents an optimization algorithm to
locally maximize this posterior distribution over $Z$
to yield a MAP estimate. Section \ref{subsecMCEM} presents
an MCEM method to select $\pi_Q$ and $\pi_\Theta$, following the empirical
Bayesian principle of maximum marginal likelihood.
An efficient implementation of all inference algorithms
is available on the authors' websites.

We emphasize that even though the BASIC model is specified hierarchically, our
inference algorithms directly sample from and maximize over the posterior
distribution of only $Z$, analytically marginalizing over the
other latent variables $q$ and $\theta$. Furthermore, these
procedures use dynamic programming subroutines that exactly sample from and
maximize over the joint conditional distribution of many or all variables in a
single row or column of $Z$, i.e.\ changepoints in a single sequence or at a
single location across all sequences. We verify in \appref{gibbscomparison}
that this greatly improves
mixing of the sampler over a na\"ive Gibbs sampling scheme that individually
samples each $Z_{j,t}$ from its univariate conditional distribution.

\subsection{Sampling from the posterior distribution}\label{subsecsampling}
To sample from $\Pr(Z|X)$, we propose the following high-level MCMC procedure:
\begin{enumerate}
\item For $j=1,\ldots,J$: Re-sample $Z_{j,\cdot}$ from
$\Pr(Z_{j,\cdot}|X,Z_{(-j),\cdot})$
\item For $t=2,\ldots,T$: Re-sample $Z_{\cdot,t}$ from
$\Pr(Z_{\cdot,t}|X,Z_{\cdot,(-t)})$
\item For $b=1,\ldots,B$: Randomly select $t$ such that $Z_{j,t}=1$ for
at least one $j$, choose $s=t-1$ or $s=t+1$, and perform a
Metropolis-Hastings step to swap $Z_{\cdot,t}$ and $Z_{\cdot,s}$.
\end{enumerate}
We treat the combination of steps 1--3 above as one complete iteration of
our MCMC sampler.
Here, $Z_{j,\cdot}$, $Z_{(-j),\cdot}$, $Z_{\cdot,t}$, and $Z_{\cdot,(-t)}$
respectively denote the $j^\text{th}$ row, all but the $j^\text{th}$ row, the 
$t^\text{th}$ column, and all but the $t^\text{th}$ column of $Z$.
In step 3, $B$ is the number of swap attempts, which we set in practice as
$B=10T$.

To sample $Z_{j,\cdot} \mid Z_{(-j),\cdot}$ in step 1, we adapt the
dynamic programming recursions developed in \citep{fearnhead} to our setting,
which require $O(T^2)$ time for each $j$. To sample $Z_{\cdot,t} \mid
Z_{\cdot,(-t)}$ in step 2, we develop a novel dynamic programming recursion
which performs this sampling in $O(J^2)$ time for each $t$. Step 3 is included
to improve the positional accuracy of detected changepoints, and the swapping of
columns of $Z$ typically amounts to shifting all changepoints at position $t$
to a new position $t+1$ or $t-1$ that previously had no changepoints. This step
may be performed in $O(JT)$ time (when $B=O(T)$), so one complete iteration of 
steps 1--3 may be performed in time $O(JT^2+J^2T)$. Details of all three
algorithmic procedures are provided in \appref{samplingalgorithms}.

\subsection{Maximizing the posterior distribution}\label{subsecmaximization}
\noindent 
To maximize $\Pr(Z|X)$ over $Z$, we similarly
propose iterating the following three high-level steps:
\begin{enumerate}
\item For $j=1,\ldots,J$: Maximize $\Pr(Z|X)$ over $Z_{j,\cdot}$.
\item For $t=2,\ldots,T$: Maximize $\Pr(Z|X)$ over $Z_{\cdot,t}$.
\item For each $t$ such that $Z_{j,t}=1$ for at least one $j$, swap
$Z_{\cdot,t}$ with $Z_{\cdot,t-1}$ or $Z_{\cdot,t+1}$ if this increases
$\Pr(Z|X)$, and repeat.
\end{enumerate}
We terminate the procedure when one iteration
of all three steps leaves $Z$ unchanged. In applications, we first perform MCMC
sampling to select $\pi_Q$ and $\pi_\Theta$ using the MCEM procedure to be
described in Section \ref{subsecMCEM}, and then initialize $Z$ in the above
algorithm to a rounded average of the sampled values. Under this
initialization, we find empirically that the above algorithm converges
in very few iterations.

To maximize $\Pr(Z|X)$ over $Z_{j,\cdot}$ in step 1, we adapt the dynamic
programming recursions developed in \citep{jacksonetal} to our setting, which
require $O(T^2)$ time for each $j$. Maximization over $Z_{\cdot,t}$ in
step 2 is easy to perform in $O(J \log J)$ time for each $t$.
Step 3 is again included
to improve the positional accuracy of detected changepoints, and after an
$O(JT)$ initialization, each swap of step 3 may be performed in $O(J)$ time.
Hence one complete iteration of steps 1--3 may be performed in time
$O(JT\log J+JT^2)$. Details of all three algorithmic procedures are provided in
\appref{maximizationalgorithms}.

\subsection{Reduction to linear cost in $T$}\label{seclineartime}
In practice, $T$ may be large, and it is desirable to improve upon the quadratic
computational cost in $T$. For sampling, one may use the particle filter
approach of \citep{fearnheadliu} in place of the exact sampling procedure in
step 1, adding a Metropolis-Hastings rejection step in the
particle-MCMC framework of \citep{andrieuetal} to correct for the approximation
error. For maximization, one may use the PELT idea of \citep{killicketal}
to prune the computation in step 1, with modifications for
a position-dependent cost as described in \citep{fanetal}.

In our applications we adopt a simpler approach of
dividing each row $Z_{j,\cdot}$ into contiguous blocks and sampling or
maximizing over the blocks sequentially; details of this algorithmic
modification are provided in
\appsref{samplingalgorithms}--\appssref{maximizationalgorithms}.
This reduces the computational cost of one iteration of MCMC sampling to
$O(J^2T)$ and of one iteration of posterior maximization to $O(JT\log J)$,
provided the block sizes are $O(1)$. In all of our simulated and real 
data examples, we use a block size of 50 data points per sequence. We
examine the effect of block size choice in \appref{gibbscomparison}.

\subsection{Empirical Bayes selection of priors $\pi_Q$ and $\pi_\Theta$}
\label{subsecMCEM}
To select $\pi_Q$ and $\pi_\Theta$ automatically using the empirical Bayes
principle of maximum marginal likelihood, we assume $\pi_Q$ is a mixture as in
\eqnref{piq} over a fixed
dictionary $\{\nu_k\}$, and we estimate the weights $\{w_k\}$.
We also assume that $\pi_\Theta$ is parametrized by a low-dimensional parameter
$\eta$, and we estimate $\eta$. We denote
$P_j(t,s)$ in \eqnref{P} by $P_j(t,s|\eta)$.

Let $\mathcal{S}(Z_{j,\cdot})$ denote the data segments
$\{(1,t_1),(t_1,t_2),\ldots,(t_k,T+1)\}$ induced by changepoints $Z_{j,\cdot}$,
i.e., $Z_{j,t_1}=\ldots=Z_{j,t_k}=1$ and $Z_{j,t}=0$ for all other $t$.
Let $N_l=\#\{t \geq 2:\sum_{j=1}^J Z_{j,t}=l\}$ be the total number of positions
where exactly $l$ sequences have a changepoint.
Our MCEM approach to maximizing the marginal likelihood over candidate priors
operates on the ``complete'' marginal log-likelihood,
\begin{align*}
&\log \Pr(X,Z|\{w_k\},\eta)\\
&=\log \Pr(X|Z,\eta)+\log \Pr(Z|\{w_k\})\\
&=\left(\sum_{j=1}^J \sum_{(t,s) \in
\mathcal{S}(Z_{j,\cdot})} \log P_j(t,s|\eta)\right)+
\sum_{l=0}^J N_l \log\left(\sum_{k \in S}
w_k\int q^l(1-q)^{J-l} \nu_k(dq)\right).
\end{align*}
Starting with the initializations $\{w_k^{(0)}\}$ and
$\eta^{(0)}$, EM iteratively computes the expected complete marginal log-likelihood (E-step)
\[l^{(i)}(\{w_k\},\eta)
=\E_{Z|X,\{w_k^{(i-1)}\},\eta^{(i-1)}}[\log \Pr(X,Z|\{w_k\},\eta)]\]
and maximizes this quantity to select new prior estimates (M-step)
\[\{w_k^{(i)}\},\eta^{(i)}=\argmax_{\{w_k\},\eta}
l^{(i)}(\{w_k\},\eta).\]
MCEM approximates the E-step by a Monte Carlo sample average,
\[\E_{Z|X,\{w_k^{(i-1)}\},\eta^{(i-1)}}[\log \Pr(X,Z|\{w_k\},\eta)]
\approx \frac{1}{M} \sum_{m=1}^M \log \Pr(X,Z^{(m)}|\{w_k\},\eta),\]
where $Z^{(1)},\ldots,Z^{(M)}$ are MCMC samples under the prior estimates
$\{w_k^{(i-1)}\}$ and $\eta^{(i-1)}$.
Maximization over $\{w_k\}$ and $\eta$ are decoupled in the M-step:
\begin{align*}
\{w^{(i)}_k\}&=\argmax_{\{w_k\}} \sum_{m=1}^M \sum_{l=0}^J N_l^{(m)}
\log\left(\sum_{k \in S} w_k\left(\int q^l(1-q)^{J-l} \nu_k(dq)\right)\right),\\
\eta^{(i)}&=\argmax_\eta \sum_{m=1}^M \sum_{j=1}^J \sum_{(t,s) \in \mathcal{S}
(Z_{j,\cdot}^{(m)})} \log P_j(t,s|\eta),
\end{align*}
where $N_l^{(m)}=\#\{t \geq 2:\sum_{j=1}^J Z_{j,t}^{(m)}=l\}$. Maximization over
$\{w_k\}$ is convex, and we use a tailored KL-divergence-minimization
algorithm for this purpose. We use a generic optimization routine to maximize
over the low-dimensional parameter $\eta$. In our applications, we take
$\{\nu_k\}_{k \in S}$ to be point masses at a grid
of points with spacing $1/J$ and spanning the range $[0,J/2)$, and we
initialize $\{w_k^{(0)}\}$ to assign large weight at 0 and spread the remaining
weight uniformly over the other grid points. We initialize $\eta^{(0)}$ by
dividing the data sequences into blocks and
matching moments. Details of the optimization and initialization procedures are 
given in \appref{MCEMalgorithms}. 

\section{Simulation studies}\label{sec:simulation}
\subsection{Assessing inference on a small example}
\label{subsecsmallexample}
\begin{figure}[tbp]
\centering
\includegraphics[width=\textwidth]{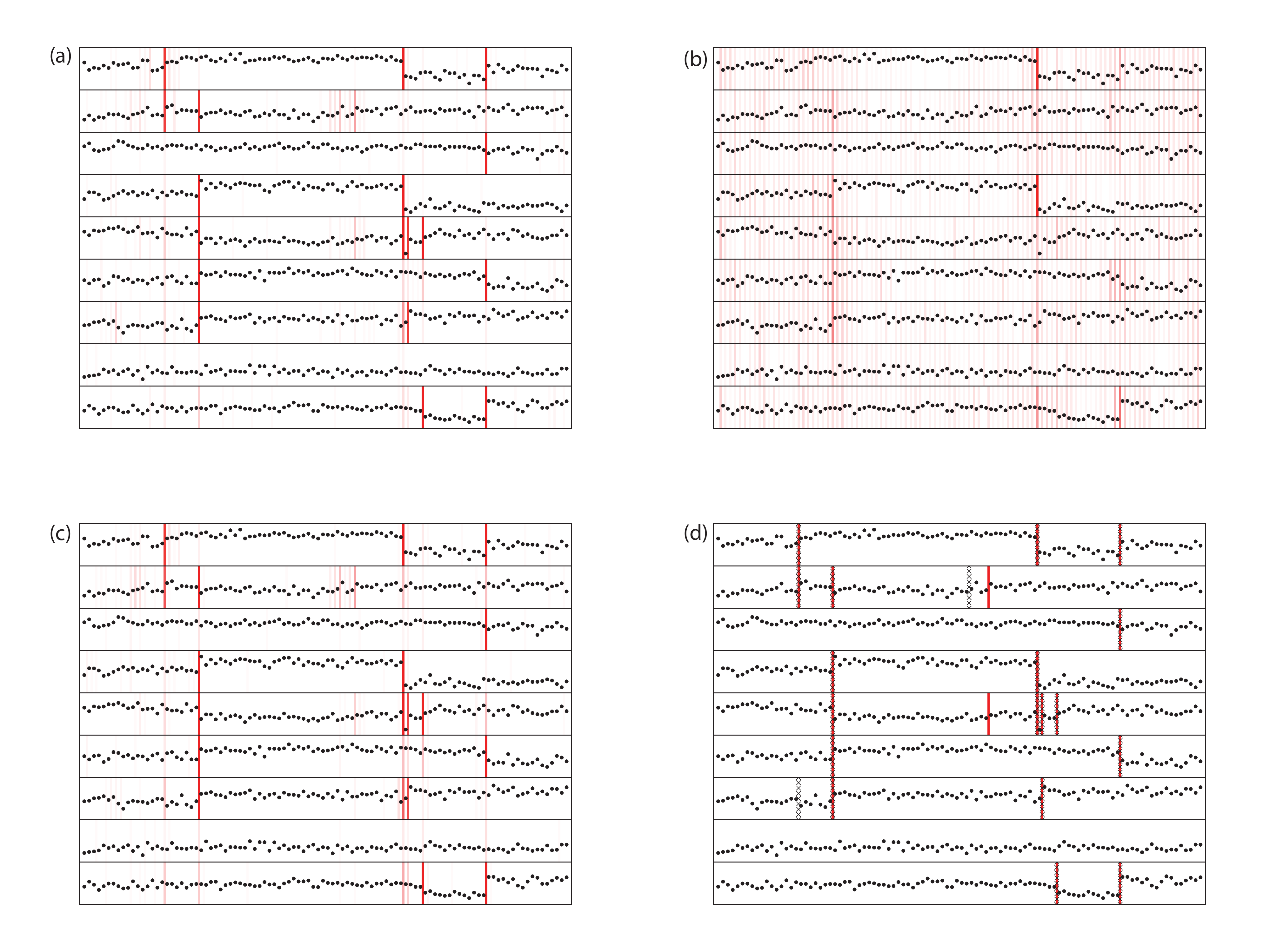}
\caption{BASIC posterior inference on data generated from the BASIC
model (see Section~\ref{subsecsmallexample}). Heatmaps (a-c) display the marginal posterior probabilities of change
$\Pr(Z_{j,t}=1|X)$ estimated by MCMC using (a) the true data-generating priors
$\pi_Q$ and $\pi_\Theta$ (which in practice are unknown), (b) grossly incorrect priors,
and (c) MCEM-selected priors. The MCEM procedure in (c) is initialized with the incorrect priors of (b) but recovers accuracy comparable to the idealized setting in (a).
Under the MCEM priors of (c), panel (d) displays the MAP changepoint estimate in red
and the true changepoints as black crosses.}\label{fig:smallexample}
\end{figure}
We first illustrate our inference procedures on the small
data example shown in Figure \ref{figmodel}, with $J=9$ sequences and
$T=100$ data points per sequence. This data was generated according to the
BASIC model (with $\theta:=(\mu,\sigma^2)$,
$p(\cdot|\theta)=\operatorname{Normal}(\mu,\sigma^2)$, $\pi_\Theta$ given by
$\mu \sim \operatorname{Normal}(0,5)$ and $\sigma^2 = 1$,
and $\pi_Q=0.9\delta_0+0.1\delta_{2/9}$). 

Figure \ref{fig:smallexample} shows the effectiveness of the empirical Bayesian 
MCEM approach to inference in this setting. Panel (a) shows the marginal posterior changepoint probabilities
$\Pr(Z_{j,t}=1|X)$ computed with 50 MCMC samples after a 50-sample burn-in,
in an idealized setting where the sampling is performed under
the true priors $\pi_Q$ and $\pi_\Theta$ that generated the data.
The results of panel (a) represent an idealized gold standard, 
as ``true priors'' are typically unknown in practice.
Panel (c) demonstrates, however, that performance comparable to the gold
standard can be obtained using MCEM-selected priors, even 
when the MCEM algorithm is initialized with a grossly incorrect prior guess.
In particular, panel (b) displays $\Pr(Z_{j,t}=1|X)$
under the grossly incorrect prior choices $\mu \sim \N(0,10)$, $\sigma^2 = 10$, and
$\pi_Q=0.2\delta_0+0.2\delta_{1/9}
+0.2\delta_{2/9}+0.2\delta_{3/9}+0.2\delta_{4/9}$,
while panel (c) displays $\Pr(Z_{j,t}=1|X)$ when prior parameters are initialized to the same grossly incorrect choices 
and updated with an MCEM update after iterations 5, 10, 20, 30, and 50 of the
burn-in. 
Notably, the posterior inferences using MCEM priors (panel (c)) are comparable to those of the idealized
setting (panel (a)), despite this incorrect initialization.
Finally, panel (d) shows the MAP estimate of $Z$ using the priors estimated
in panel (c). In this example, the MAP estimate misses two true changepoints
and makes two spurious detections.

We repeated this simulation with 100 different data sets generated from
the BASIC model. Table \ref{tablesmallexample} summarizes results using
three error measures (all averaged across the 100 experiments):
the squared error of the posterior mean changepoint indicators
$\sum_{j,t} (\E\{Z_{j,t}\mid X\} -Z_{j,t}^\true)^2$,
the squared error of the posterior mean signal reconstruction $\sum_{j,t} (\E\{\theta_{j,t}\mid X\}-\theta_{j,t}^\text{true})^2$, 
and the 0--1 error of detected changepoints in the MAP estimate.
All evaluation metrics indicate that posterior inference using the
MCEM-selected prior consistently leads to accuracy comparable 
to the idealized gold standard of using the true data-generating prior.
As a reference
point for the difficulty of this simulated data, the average 0--1
changepoint error of applying a univariate changepoint method
(PELT with default MBIC penalty in the ``changepoint'' R package,
\cite{killicketal}) to each data sequence
individually is 12.6, which is 25\% higher than that of our MAP estimate under
the MCEM-selected prior.

\begin{table}[t]
\caption{Errors averaged over 100 instances of the
Section~\ref{subsecsmallexample} simulation. Posterior inference using
MCEM-selected priors recovers accuracy comparable to the idealized setting of
using the true data-generating priors (``True priors''), even when initialized with grossly incorrect priors (``Wrong priors'').}
\label{tablesmallexample}
\begin{tabular}{lccc}\toprule
& \textbf{True priors} & \textbf{Wrong priors} & \textbf{MCEM priors}\\
\midrule
Squared error of $\E\{Z\mid X\}$ & 8.1 & 17.9 & 8.3\\
Squared error of $\E\{\theta\mid X\}$ & 50.3 & 151 & 51.1\\
0--1 changepoint error of $Z^\text{MAP}$ & 10.3 & 14.9 & 10.1\\
\bottomrule
\end{tabular}
\end{table}

\subsection{Comparing detection accuracy on artificial CNV data}
\label{subsecsyntheticCNV}
The identification of copy number variations (CNVs) in aCGH data for cancer
cells represents one primary motivation for our work. As there is typically no
known ``gold standard'' for the locations of all CNVs in real aCGH data, we will
assess changepoint detection accuracy in a simulation study, applying our
inference procedures to 50 simulated aCGH data sequences using the simulator
from
\cite{louhimoetal}\footnote{This simulator also generates corresponding gene
expression data; we ignored this additional data, as integration of these two
data types is not the focus of our paper.}. This simulator generates
six CNVs that are either focal high-level (2-copy loss or 6-to-8-copy gain),
focal medium-level (1-copy loss or 4-copy gain),
or broad low-level (1-copy gain).
The prevalence of each CNV across samples ranges between 5\%
and 50\%. The simulator accounts for sample
heterogeneity, with each sample corresponding to a random mixture of normal
and abnormal cells.

\begin{figure}[t]
\begin{minipage}{0.6\textwidth}
\includegraphics[width=\textwidth]{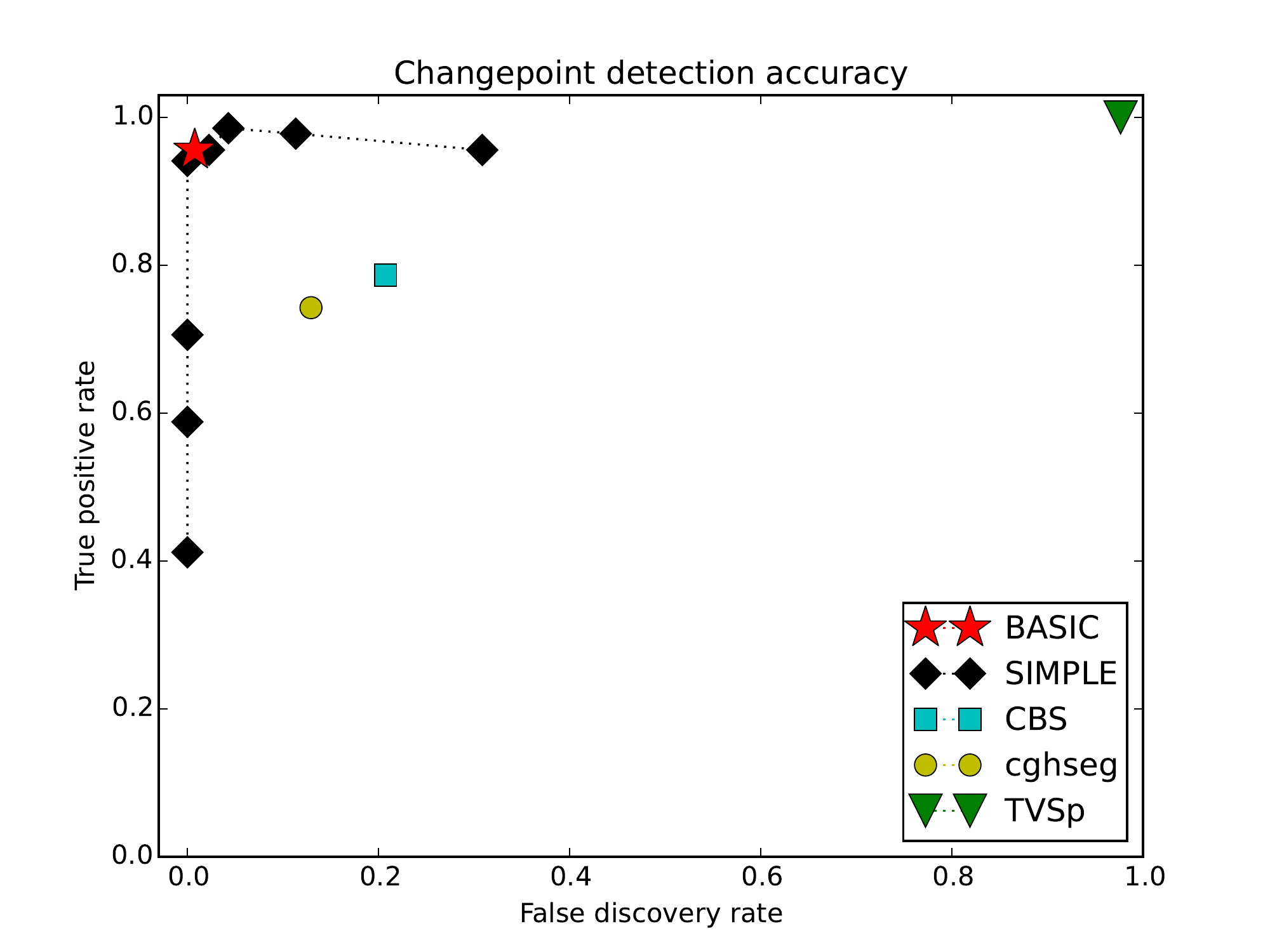}
\end{minipage}%
\begin{minipage}{0.4\textwidth}
\begin{center}
Signal reconstruction error\\
\vspace{0.1in}
\begin{tabular}{c|c}
Method & $\sum_{j,t} (\mu_{j,t}^\text{est}-\mu_{j,t}^\text{true})^2$\\
\hline
BASIC & 10.40 \\
SIMPLE & 10.42 \\
CBS & 21.82 \\
cghseg & 29.23 \\
TVSp & 54.22
\end{tabular}
\end{center}
\end{minipage}
\caption{Changepoint detection accuracy and signal reconstruction squared-error
for various methods on simulated aCGH data from \cite{louhimoetal} (see Section
\ref{subsecsyntheticCNV}). Left:
Fraction of true changepoints detected across all sequences, versus fraction of
all changepoint detections that are false discoveries. Right: Total signal
reconstruction squared-error, where $\mu_{j,t}^\text{est}$ is the estimated
$\log_2$ ratio at probe $t$ in sequence $j$, and $\mu_{j,t}^\text{true}$ is its
true value. For SIMPLE, we report the highest accuracy obtained across all values of its tuning parameter.}\label{figsyntheticCNV}
\end{figure}

To apply BASIC, we performed 100 iterations of MCMC sampling after 100
iterations of burn-in, using a normal likelihood model with changing mean and
fixed (unknown) variance, and with MCEM updates of prior
parameters after iterations 10, 20, 40, 60, and 100 of the burn-in.
We then performed MAP estimation using the resulting empirical Bayes priors,
with $Z$ initialized to the MCMC sample average. On this data, the BASIC MAP estimate
achieved 100\% accuracy; we report results in \appref{syntheticCNV}.

One way in which this synthetic data is easier than the real aCGH data
we analyze in Section \ref{secCNV} is that focal and broad
CNVs span at least 50 and 500 probes, respectively, whereas they are shorter in
our data of Section \ref{secCNV} and also in certain previous single-sample
comparison studies \citep{laietal}. To increase the difficulty in this regard,
we subsampled every tenth point of each synthetic data sequence and analyzed the
resulting sequences, in which focal CNVs span 5 probes and broad CNVs span 50.
Results on this more challenging dataset are reported here.

The accuracy of the BASIC MAP estimate is shown as the red star in Figure
\ref{figsyntheticCNV}, where we plot the fraction of true
changepoints discovered against the false-discovery proportion.
Shown also in Figure \ref{figsyntheticCNV} are the results of
several alternative methods: SIMPLE \citep{fanetal} to represent the penalized
likelihood approach, TVSp \citep{zhouetal} to represent total-variation
regularization, circular binary segmentation (CBS)
\citep{olshenetal} applied separately to each sequence to represent a popular
method of unpooled analysis, 
and cghseg \citep{picardetal} to represent a popular method of pooled analysis.
We set the convergence
tolerance of TVSp to $10^{-14}$ and ignored changes with mean shift
less than $10^{-3}$ to avoid identifying breakpoints because of numerical
inaccuracy. We applied
SIMPLE with a normal likelihood model; as the method does not prescribe a
default value for the main tuning parameter, we plot its performance as the
tuning parameter varies. All remaining parameters of the methods were set to
their default values or selected using the provided cross-validation routines.

Detection accuracy of the BASIC MAP estimate is near-perfect and competitive
with the other tested methods---examination of its output reveals that it
misses a focal (5-probe) medium-level loss in two sequences and a broad
low-level gain in one sequence, and it makes one spurious segmentation in one
sequence. Detection by cghseg is conservative, missing 10 focal gains and losses
across all sequences. In addition, as cghseg does not attempt to identify
changepoints at common sequential positions, it inaccurately identifies the
location of 15 additional changepoints, which contributes both to an increased
false discovery proportion and a reduced true discovery proportion. (This
positional inaccuracy ranges between 1 and 5 probes.) Single-sequence CBS
suffers from
the same changepoint location inaccuracy. It is less conservative than cghseg,
truly missing only 3 aberrations across all sequences,
but also identifying 2 non-existent aberrations.
TVSp partitions the data sequence
into too many segments, yielding false-discovery proportion close to 1 for
changepoint discovery. We do note that TVSp and its tuning-parameter selection
procedure are designed to minimize the signal-reconstruction squared error,
rather than changepoint identification error. However, we report the signal
reconstruction errors alongside Figure \ref{figsyntheticCNV} and observe that
TVSp is also less accurate by this metric.

SIMPLE yields performance close to that of BASIC under optimal tuning, but the
authors of \citep{fanetal} provide little guidance on how to choose the tuning
parameter. In the BASIC framework, the analogous hyperparameters of
$\pi_Q$ are selected automatically by MCEM.

\section{Copy number aberrations in the NCI-60 cancer cell lines}
\label{secCNV}
We applied our BASIC model to analyze CNVs
in aCGH data for the NCI-60 cell lines, a set of 60 cancer cell
lines derived from human tumors in a variety of tissues and organs, as reported
in \citep{varmaetal}. We discarded measurements on the sex chromosomes,
removed outlier measurements, and centered each sequence to have median 0; we
discuss these preprocessing steps in \appref{NCI60}. We fit the BASIC model
using a normal likelihood with changing mean and fixed variance, applying the
same procedure as in Section \ref{subsecsyntheticCNV}. The runtime of our
analysis on the pooled data ($J=125,T=40217$) was 2 hours.

\begin{figure}
\hspace*{-0.3in}\includegraphics[width=0.6\textwidth]{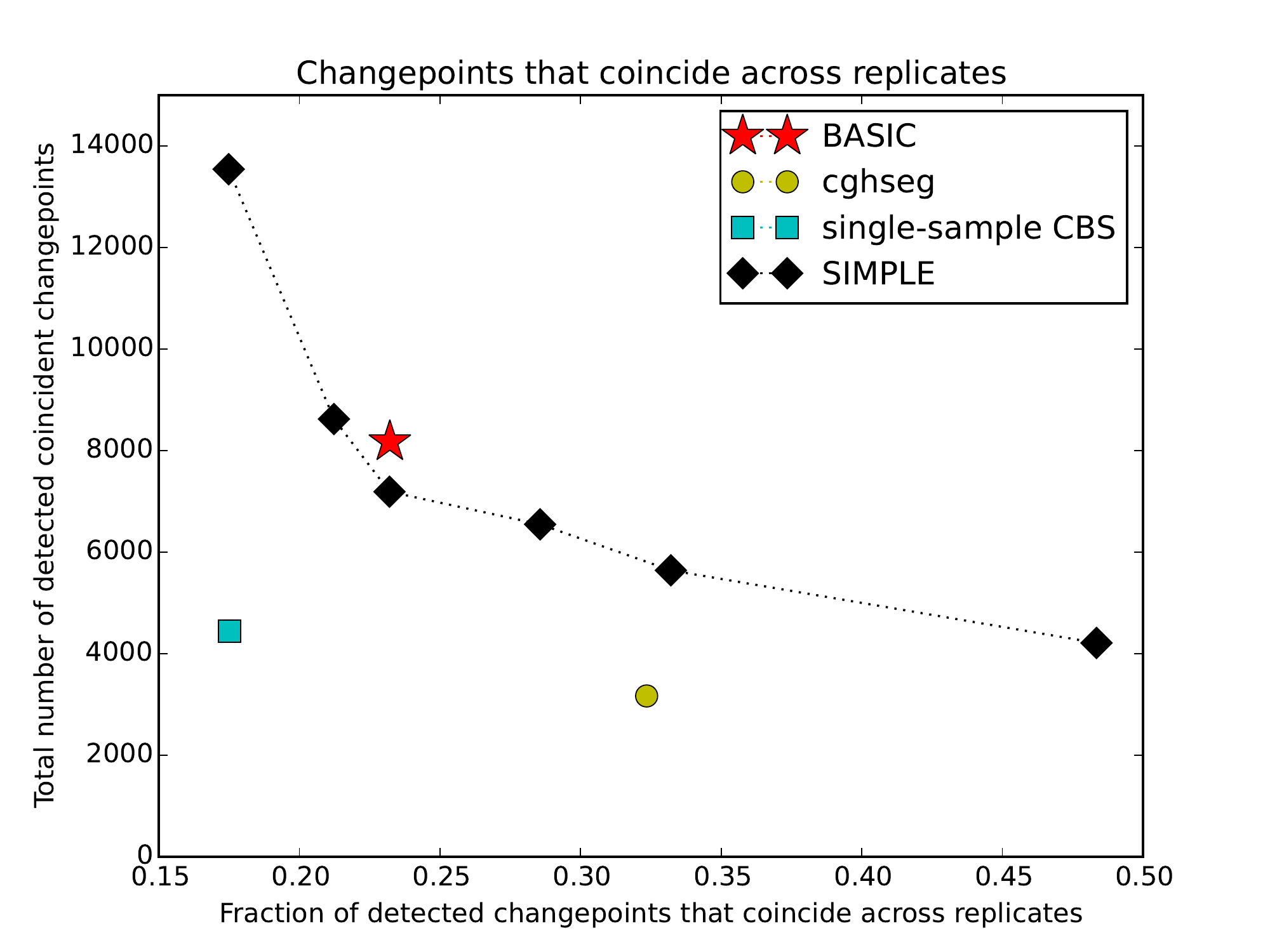}
\caption{Comparison of methods by the total number and rate of detected
changepoints that are coincident across two technical replicates of real aCGH
data for 59 cancer cell lines (see Section \ref{secCNV}). The performance of SIMPLE varies with its unspecified tuning parameter.}\label{figtwosample}
\end{figure}

In this data, measurements for 59 of the 60 cell lines were made with at least
two technical replicates. We used this to test the changepoint detection
consistency of various methods, by constructing two data sets of
59 sequences corresponding to the two replicates and applying each method
to the data sets independently. A detected changepoint is ``coincident'' across
replicates if it is also detected in the same cell line at the same probe
location in the other replicate. Figure \ref{figtwosample} plots the total
number of coincident detections versus the fraction of all changepoint
detections that are coincident, for the methods tested in Section
\ref{subsecsyntheticCNV}. (We omit the comparison with TVSp due to its high
false-discovery rate for changepoint identification.)
BASIC has better performance than single-sample CBS, yielding more
coincident detections also at a higher coincidence rate. BASIC
is less conservative than cghseg, detecting more coincident changes but at a
lower coincidence rate. 
Recall that the performance of SIMPLE varies with its unspecified tuning parameter.
For comparable tunings of SIMPLE, BASIC yields slightly better performance:
for the same level of changepoint coincidence across replicates, BASIC detects
more changepoints, and for comparable numbers of detected changepoints, BASIC
achieves a higher level of changepoint coincidence.

We emphasize that a non-coincident detection is not
necessarily wrong---for a changepoint demarcating a low-level aberration against
which a method does not have full detection power, a method may
detect this change in one replicate but not the other. Conversely, a
coincident detection need not correspond to a true CNV, if technical
artifacts are present in both replicates. The coincidence rate is
not high for any tested method. Reasons for this include (1) changepoints due
to technical drift,
a common occurrence~\citep{olshenetal} which is particularly severe in some of the
sequences of this data set; (2) probe artifacts that differ across replicates;
and (3) low-level non-shared aberrations with boundary points that are
difficult to precisely identify. The coincidence rate may be
increased for all methods by applying post-processing procedures to remove
changepoints due to technical drift and probe artifacts, although these
procedures are usually ad hoc.

\begin{figure}[tbp]
\hspace*{-0.3in}\includegraphics[width=\textwidth]{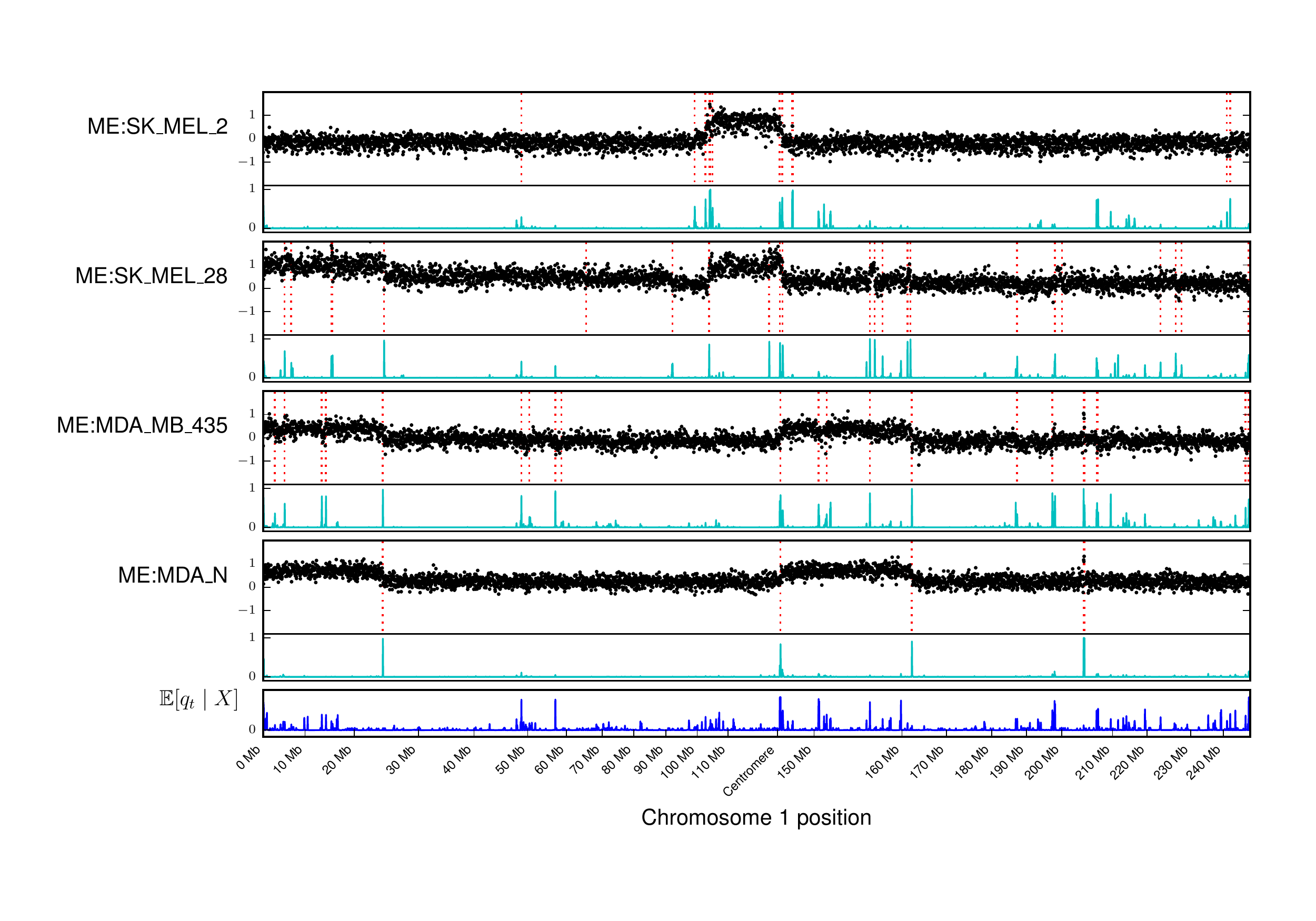}
\caption{Chromosome 1 aCGH measurements for four NCI-60 melanoma cell lines (black points)
and associated  BASIC estimates of marginal posterior changepoint probabilities
using 100 MCMC samples (teal curves).
Red dashed lines indicate BASIC MAP changepoint
estimates. The estimated posterior mean of $q_t$ is displayed below in blue, providing a
cross-sample summary of changepoint prevalence across all 125 analyzed
sequences.}\label{figmelanoma}
\end{figure}

Our BASIC framework provides not only a point estimate of changepoints, but also
posterior probability estimates that may be valuable in interpreting results and
also performing this type of post-processing.
Figure \ref{figmelanoma} displays the $\log_2$-ratio measurements and the
BASIC MAP estimate of changepoints in chromosome 1 for four
distinct melanoma cell lines,
alongside the estimated marginal posterior changepoint
probabilities. Figure \ref{figmelanoma}
also displays the posterior mean estimate of $q_t$ (computed
from the sampled $Z$ matrices), which provides a cross-sample summary
of the prevalence of shared changepoints across all analyzed sequences at each
probe location.

\begin{figure}[tbp]
\includegraphics[width=0.9\textwidth]{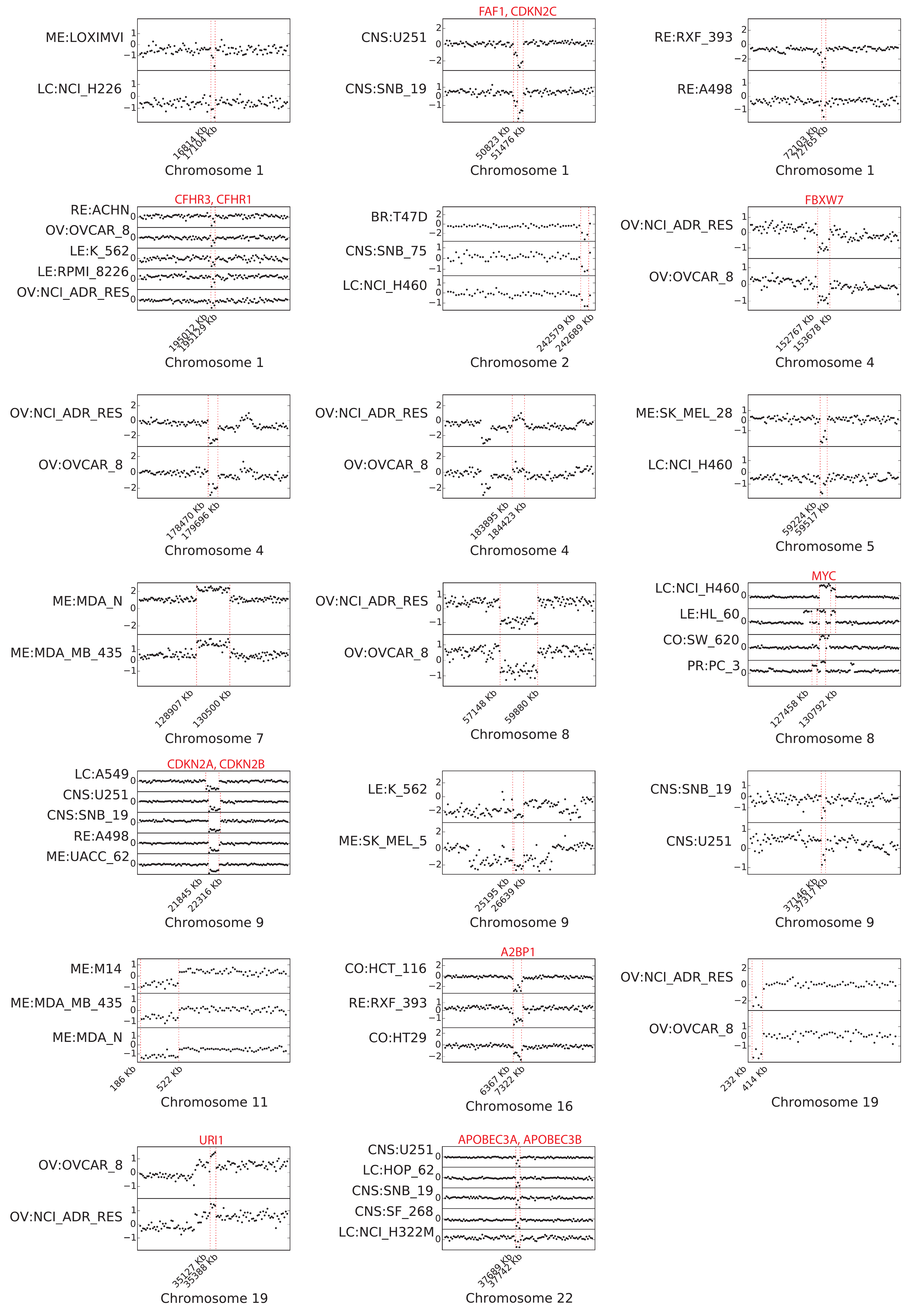}
\caption{The 20 most prominent focal CNVs present in at least two of the
NCI-60 cancer cell lines. Genes of interest in the aberrant regions are
highlighted in red.}
\label{figshortCNVs}
\end{figure}

To illustrate one use of this posterior information, we performed a
pooled analysis of all sequences (including all replicates to increase detection
power and accuracy) in order to highlight genomic locations that
contain focal and shared CNVs.
First, we identified all pairs of genomic locations $s$ and $t$ on the 
same chromosome at distance less than $3 \times 10^6$ base
pairs apart\footnote{We use 3 million base pairs as the cut-off to distinguish
focal from non-focal CNVs.}
such that at least two distinct cell lines had 
posterior probability greater than 90\% of containing changepoints at both $s$
and $t$. The interval between $s$ and $t$
is the identified CNV, and the sequences having posterior probability greater
than 90\% of change at $s$ and $t$ are the identified carriers of that CNV. To
reduce false discoveries due to technical noise of the
aCGH experiments, we restricted attention to those pairs for which
this interval contained at least three microarray probes. Then, for each such
pair, we computed the mean value of the data in the
interval between $s$ and $t$ for the carrier sequences and compared this to
the mean value in small intervals before $s$ and after $t$.
Figure \ref{figshortCNVs} shows the 20 identified CNVs that
exhibit the greatest absolute difference between these mean values, displaying
up to five distinct carriers of each CNV. CNVs that overlap in
genomic position are grouped together in the figure.

Many of the CNVs highlighted in Figure \ref{figshortCNVs} contain genes that
have been previously studied in relation to cancer; we have annotated the
figure with some of these gene names. CDKN2A and CDKN2B are
well-known tumor suppressor genes whose deletion and mutation have been observed
across many cancer types \citep{kambetal, noborietal}. FBXW7 is
a known tumor suppressor gene that plays a role in cellular division
\citep{akhoondietal}. MYC is a well-known oncogene that is commonly amplified in
many cancers \citep{dang}. URI1 is a known oncogene in ovarian cancer
\citep{theurillatetal}. FAF1 is believed to be a tumor suppressor gene involved
in the regulation of apoptosis \citep{mengesetal}. Deletion of A2BP1
has been previously observed in colon cancer tumors and gastric cancer cell
lines \citep{trautmannetal, tadaetal}. Deletion of APOBEC3
has been observed in breast cancer \citep{longetal, xuanetal}, although
we detect its deletion in cell lines of cancers of the central nervous system
and the lung. Deletion of CFHR3 and CFHR1 is not specifically
linked to cancer, but it is a common haplotype that has been observed in many
healthy individuals \citep{hughesetal}. Many of the remaining CNVs in Figure
\ref{figshortCNVs} appear to represent true copy number variations present
in the data (rather than spurious detections by our algorithm), but we could
not validate the genes present in the corresponding genomic regions against the
cancer genomics literature.

\section{Price volatility in S\&P 500 stocks}
\label{secstocks}
\begin{figure}[t]
\includegraphics[width=\textwidth]{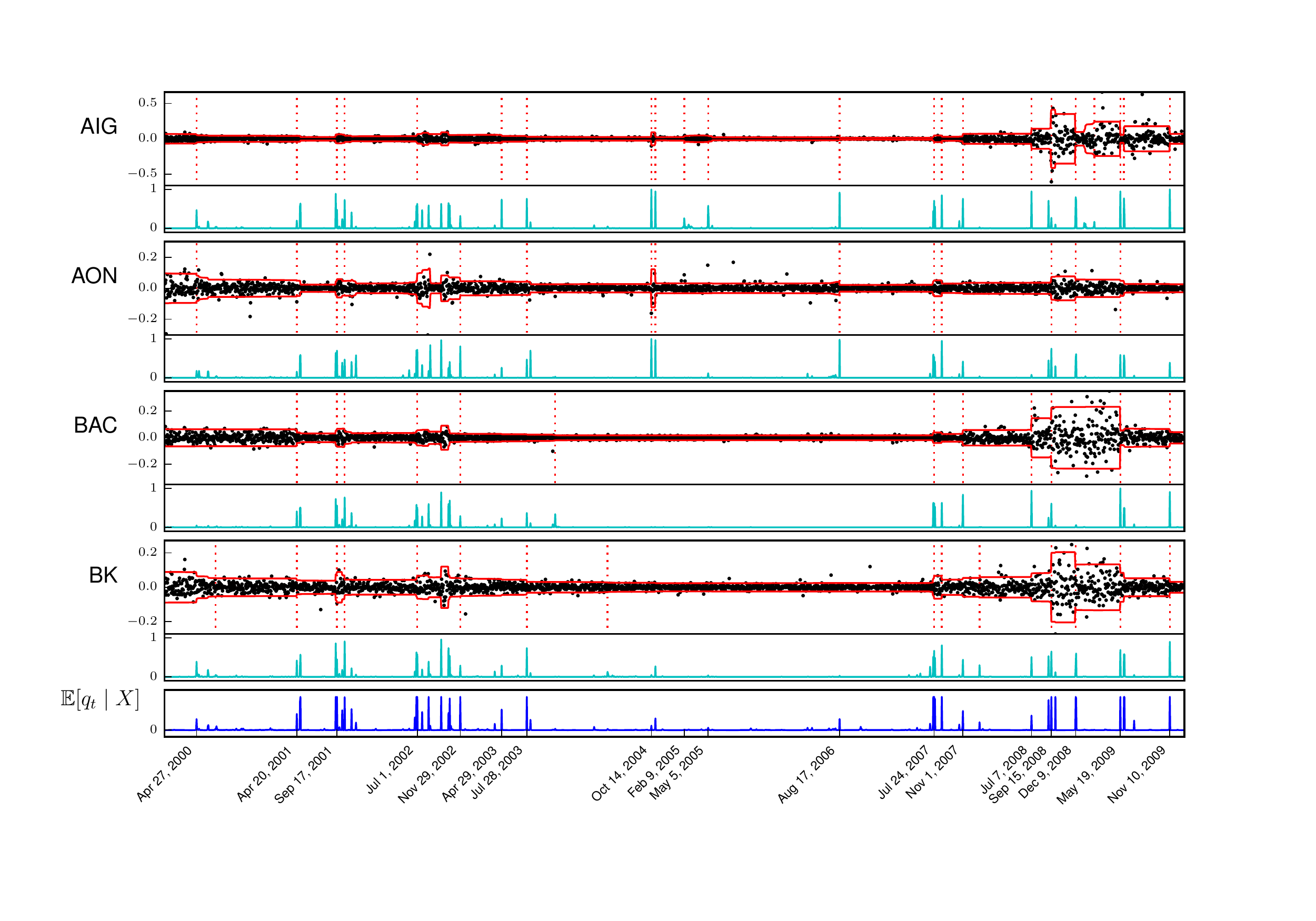}
\caption{Daily returns of four U.S.\ stocks from 2000 to 2009, with MAP
changepoint estimates (from a joint analysis of 401 stocks)
shown in dashed red and model-based volatility
estimates shown in solid red. The estimated posterior mean of $q_t$ is displayed
below in blue.}\label{figstocks}
\end{figure}

As a second example, we applied the BASIC model to
analyze the volatility in returns of U.S. stocks from the year 2000 to 2009. We
collected from Yahoo Finance the daily adjusted closing prices of stocks that
were in the S\&P 500 index fund over the entire duration of this 10-year
period, and we computed the daily return of each stock on each trading day $t$
as $(p_t-p_{t-1})/p_{t-1}$, where $p_t$ is its closing price on day $t$
and $p_{t-1}$ is its closing price on the previous day. Our data
consists of the returns for $J=401$ stocks over $T=2514$ trading days, and the
total runtime of our pooled analysis was 1 hour.

Previous authors have applied univariate changepoint detection methods to
analyze daily returns of the Dow Jones Industrial Index
from 1970 to 1972, modeling the data as normally distributed with zero mean and
piecewise constant variance \citep{hsu,adamsmackay}. We observed empirically for
our data that the tails of the distribution of daily returns are heavier
than normal, and we instead applied BASIC using a Laplace likelihood with fixed
zero mean and piecewise constant scale. We used the same MCMC/MCEM/MAP inference
procedure as in Section \ref{subsecsyntheticCNV}.
 
Shown in Figure \ref{figstocks} are the daily returns for American International
Group Inc.\ (AIG), Aon Corp.\ (AON), Bank of America Corp.\ (BAC), and
The Bank of New York Mellon Corp.\ (BK), together with MAP
changepoint estimates and estimated marginal posterior change probabilities.
Shown also is the cross-sample changepoint summary provided by the posterior
mean of $q_t$. Within this 10-year period,
the 15 trading days with the highest posterior mean for $q_t$ are, in
chronological order: Sep 6 2001, Sep 17 2001, Jun 27 2002, Jul 1 2002, Aug 9
2002, Sept 24 2002, Nov 29 2002, Jul 24 2007, Aug 20 2007, 
Sep 15 2008, Sep 29 2008, Dec 9 2008, Jun 2 2009, Jun 3 2009, and
Nov 10 2009. The
changepoints from 2001 to 2002 are attributable to the collapse of
the dot-com bubble of the late 1990s and early 2000s, and those from 2007
to 2009 are attributable to the U.S.\ financial crisis. Several of these
dates correspond to important events in U.S.\ stock market history,
including Sep 17 2001 when the markets first re-opened after the
World Trade Center terrorist attacks, Jul 1 2002 when
WorldCom stock fell in value by 93\%, Sept 15 2008 when
Lehman Brothers filed for Chapter 11 bankruptcy, and Sept 29 2008 
when the U.S.\ House of Representatives rejected a proposed bailout plan
for the financial crisis.

Many other detected changepoints were local to small numbers of individual
stocks. For instance, the
changepoint detected on Oct 14 2004 and visible in the first two sequences of
Figure \ref{figstocks} was shared across the seven stocks AIG, AON,
Coventry Health Care, Hartford Financial Services, Marsh
\& McLennan, Merk \& Co., and Unum Group. Six of
these seven stocks belong to the insurance industry, and the
changepoint represents a brief spike in price volatility due to an
insurance scandal that was revealed on Oct 14 2004 when AIG publicly
disclosed its involvement, along with Marsh \& McLennan and others, in an
illegal market division scheme, and civil and criminal charges were announced
against Marsh \& McLennan and employees at AIG pertaining to various
allegations of corporate misbehavior.\footnote{Source: ``Just how rotten?'',
{\it The Economist}, Special Report, 21 October 2004.} Other examples of
detected ``locally-shared'' changepoints include Oct 10 2000, marking the
beginning of a period of increased price volatility in the tech companies
Amazon.com, Cisco Systems, EMC Corporation, JSD Uniphase,
Oracle Corporation, and Yahoo! Inc.; and
Feb 16 2005, coinciding with the date on which the international Kyoto Protocol
treaty on carbon emissions took effect and marking the start of a period of
increased price volatility in the energy companies Dominion Resources,
Devon Energy, Public Service Enterprise Group, and Exxon Mobil.

We may also use our methods to produce a smooth estimate of the historical
volatility of stock prices, by computing the
posterior mean of the Laplace scale parameter $\theta_{j,t}$ for each sequence
$j$ and each day $t$ using the sampled $Z$ matrices. The Laplace scale
parameter $\theta_{j,t}$ implies a standard deviation of $\sqrt{2}\theta_{j,t}$;
red lines in Figure \ref{figstocks} are plotted at $\pm 2$
standard deviations to pictorially illustrate this volatility estimate. 
This estimate is smooth and resilient to outliers, while
still exhibiting rapid adjustments to real structural changes in the data.

\appendix
\section{Likelihood models}\label{sec:likelihoodmodels}
For concreteness, we record here several practically-relevant choices of $p(\cdot|\theta)$ and
$\pi_\Theta$ in the BASIC model, along with the corresponding computations for $P_j(t,s)$ in
\eqnref{P}. In each of these
settings, the prior distribution $\pi_\Theta$ is parametric, and we denote the
parameter of $\pi_\Theta$ as $\eta$.\\

\noindent \emph{Normal model, changing mean and fixed variance:}
\begin{align}\label{eqn:normalfixedvar}
&\theta\defeq (\mu,\sigma^2), \hspace{0.2in} X_{j,t}|\theta \sim \operatorname{Normal}
(\mu,\sigma^2)\\
\nonumber &\eta\defeq (\mu_0,\lambda,\sigma_0^2), \hspace{0.2in}
\mu|\eta \sim \operatorname{Normal}(\mu_0,\tfrac{\sigma_0^2}{\lambda}),
\hspace{0.2in} \sigma^2|\eta \equiv \sigma_0^2\\
\nonumber &P_j(t,s)
=(2\pi \sigma_0^2)^{-\frac{s-t}{2}}\sqrt{\frac{\lambda}{\lambda+s-t}}
\exp\left(-\frac{\lambda \mu_0^2+\sum_{r=t}^{s-1} X_{j,r}^2-\frac{(\lambda \mu_0
+\sum_{r=t}^{s-1} X_{j,r})^2}{\lambda+s-t}}{2\sigma_0^2}\right)
\end{align}

\noindent \emph{Normal model, changing variance and fixed mean:}
\begin{align}\label{eqn:normalfixedmean}
&\theta\defeq (\mu,\sigma^2), \hspace{0.2in} X_{j,t}|\theta \sim \operatorname{Normal}
(\mu,\sigma^2)\\
\nonumber &\eta\defeq (\mu_0,\alpha,\beta), \hspace{0.2in}
\sigma^2|\eta \sim \operatorname{InverseGamma}(\alpha,\beta),
\hspace{0.2in} \mu|\eta \equiv \mu_0\\
\nonumber &P_j(t,s)
=(2\pi)^{-\frac{s-t}{2}}\frac{\beta^\alpha}{\Gamma(\alpha)}
\frac{\Gamma\left(\alpha+\frac{s-t}{2}\right)}{\left(\beta
+\frac{(s-t)\mu_0^2}{2}+\sum_{r=t}^{s-1}
\frac{X_{j,r}^2}{2}-\mu_0\sum_{r=t}^{s-1}X_{j,r}\right)^{\alpha+\frac{s-t}{2}}}
\end{align}

\noindent \emph{Normal model, changing mean and variance:}
\begin{align}\label{eqn:normal}
&\theta\defeq (\mu,\sigma^2), \hspace{0.2in} X_{j,t}|\theta \sim \operatorname{Normal}
(\mu,\sigma^2)\\
\nonumber &\eta\defeq (\mu_0,\lambda,\alpha,\beta), \hspace{0.2in}
\sigma^2|\eta \sim \operatorname{InverseGamma}(\alpha,\beta),
\hspace{0.2in} \mu|\sigma^2,\eta \sim
\operatorname{Normal}\left(\mu_0,\tfrac{\sigma^2}{\lambda}\right)\\
\nonumber &P_j(t,s)
=\sqrt{\frac{\lambda}{\lambda+s-t}}\frac{\beta^\alpha}{\Gamma(\alpha)}
(2\pi)^{-\frac{s-t}{2}}\frac{\Gamma\left(\alpha+\frac{s-t}{2}\right)}
{\left(\beta+\frac{\lambda\mu_0^2+\sum_{r=t}^{s-1} X_{j,r}^2}{2}
-\frac{(\lambda \mu_0+\sum_{r=t}^{s-1} X_{j,r})^2}{2(\lambda+s-t)}
\right)^{\alpha+\frac{s-t}{2}}}
\end{align}

\noindent \emph{Poisson model, changing mean:}
\begin{align}\label{eqn:poisson}
&\theta\defeq \lambda, \hspace{0.2in} X_{j,t}|\theta \sim \operatorname{Poisson}(\lambda)\\
\nonumber &\eta\defeq (\alpha,\beta), \hspace{0.2in} \lambda|\eta \sim
\operatorname{Gamma}(\alpha,\beta)\\
\nonumber &P_j(t,s)
=\left(\prod_{r=t}^{s-1} \frac{1}{X_{j,r}!}\right)
\frac{\beta^\alpha}{\Gamma(\alpha)}\frac{\Gamma(\alpha+\sum_{r=t}^{s-1}
X_{j,r})}{(\beta+1)^{\alpha+\sum_{r=t}^{s-1} X_{j,r}}}
\end{align}

\noindent \emph{Bernoulli model, changing success probability:}
\begin{align}\label{eqn:bernoulli}
&\theta\defeq p, \hspace{0.2in} X_{j,t}|\theta \sim \operatorname{Bernoulli}(p)\\
\nonumber &\eta\defeq (\alpha,\beta), \hspace{0.2in} p|\eta \sim
\operatorname{Beta}(\alpha,\beta)\\
\nonumber &P_j(t,s)=\frac{\Gamma(\alpha+\beta)}
{\Gamma(\alpha)\Gamma(\beta)}\frac{\Gamma(\alpha+\sum_{r=t}^{s-1} X_{j,r})
\Gamma(\beta+s-t-\sum_{r=t}^{s-1} X_{j,r})}{\Gamma(\alpha+\beta+s-t)}
\end{align}

\noindent \emph{Laplace model, changing scale and fixed zero mean:}
\begin{align}\label{eqn:laplace}
&\theta\defeq \nu, \hspace{0.2in} X_{j,t}|\theta \sim \operatorname{Laplace}(0,\nu)\\
\nonumber &\eta\defeq (\alpha,\beta), \hspace{0.2in} \nu|\eta \sim
\operatorname{InverseGamma}(\alpha,\beta)\\
\nonumber &P_j(t,s)=2^{-(s-t)}\frac{\beta^\alpha}{\Gamma(\alpha)}
\frac{\Gamma(\alpha+s-t)}
{\left(\beta+\sum_{r=t}^{s-1} |X_{j,r}|\right)^{\alpha+s-t}}
\end{align}

\section{MCMC sampling algorithms}\label{sec:samplingalgorithms}
Below are the details of the MCMC sampling steps discussed in Section
\ref{subsecsampling}. Throughout, we define the quantities
\begin{align}
f(k)&=\int q^k(1-q)^{J-k}\pi_Q(dq),\label{eqn:fk}\\
g(k)&=\int q^{k-1}(1-q)^{J-k}\pi_Q(dq)\label{eqn:gk},
\end{align}
for $k=0,\ldots,J$ in \eqnref{fk} and $k=1,\ldots,J$ in \eqnref{gk}.
These quantities depend only on $\pi_Q$ and may be pre-computed
outside of the sampling iterations.
(If $\pi_Q$ is discrete or a mixture of Beta distributions, these quantities
are easily computed analytically. Otherwise, these may be computed numerically
for each $k$.) The computational costs of our MCMC sampling and MAP estimation
procedures depend on $\pi_Q$ only via pre-computation of $f(k)$ and $g(k)$.\\

\noindent \emph{Step 1: Gibbs sampling by rows}\\

To sample each row $Z_{j,\cdot}$ conditional on the remaining rows
$Z_{(-j),\cdot}$, we may employ the dynamic programming recursions developed by
Paul Fearnhead for the univariate changepoint problem \citep{fearnhead}, in the
following manner.

Let $N_j(t)=\left(\sum_{j'=1}^J Z_{j',t}\right)-Z_{j,t}$
denote the number of changepoints at position $t$ in all but the $j^\text{th}$
sequence, and let $\Pr^{(j)}$ denote probability conditional on $Z_{(-j),\cdot}$,
with associated conditional expectation $\E^{(j)}$. Note that $N_j(t)$ is
deterministic under $\Pr^{(j)}$. Then the probability density function of
$q_t$ conditional on $Z_{(-j),\cdot}$ is given, for each $q \in S$, by
\[\Pr^{(j)}(q_t=q) \propto
\Pr\left(Z_{(-j),t}|q_t=q\right)\Pr(q_t=q)
=q^{N_j(t)}(1-q)^{J-N_j(t)-1}\Pr(q_t=q).\]
Letting $c_j(t)\defeq\Pr^{(j)}(Z_{j,t}=1)=\E^{(j)}[q_t]$, this implies that
\begin{equation}\label{eqn:ct}
c_j(t)=\frac{f(N_j(t)+1)}{g(N_j(t)+1)}.
\end{equation}
For each $t>1$, let $Q_j(t)=\Pr^{(j)}(X_{j,t:T}|Z_{j,t}=1)$, and let
$Q_j(1)=\Pr^{(j)}(X_{j,1:T})$. $Q_j(t)$ is the joint probability density of
the observed data in sequence $j$ after and including position $t$,
conditional on a
changepoint having occurred in sequence $j$ at position $t$ and also
conditional on the observed changepoints in all of the other sequences. Let $P_j(t,s)$ be
as defined in \eqnref{P}. Then $Q_j(t)$ satisfies
the following recursions, which are similar to those in Theorem 1 of
\citep{fearnhead}:
\begin{align}
\nonumber Q_j(T)&=\Pr^{(j)}\left(X_{j,T}|Z_{j,T}=1\right)\\
&=P_j(T,T+1),\label{eqn:QT}\\
\nonumber Q_j(t)&=\Bigg(\sum_{s=t+1}^T
\Pr^{(j)}(Z_{j,(t+1):(s-1)}=0,Z_{j,s}=1|Z_{j,t}=1) \times \\
\nonumber
&\hspace{1in}\Pr^{(j)}\left(X_{j,t:T}|Z_{j,t}=1,Z_{j,(t+1):(s-1)}=0,Z_{j,s}=1
\right)\Bigg)\\
\nonumber &\hspace{0.5in}+\Pr^{(j)}(Z_{j,(t+1):T}=0|Z_{j,t}=1)
\Pr^{(j)}\left(X_{j,t:T}|Z_{j,t}=1,Z_{j,(t+1):T}=0\right)\\
\nonumber &=\Bigg(\sum_{s=t+1}^T
\left(\prod_{r=t+1}^{s-1}\Pr^{(j)}(Z_{j,r}=0)\right)\Pr^{(j)}(Z_{j,s}=1)\times\\
\nonumber &\hspace{1in}
\Pr\left(X_{j,t:(s-1)}|Z_{j,t}=1,Z_{j,(t+1):(s-1)}=0,Z_{j,s}=1\right)
\Pr^{(j)}\left(X_{j,s:T}|Z_{j,s}=1\right)\Bigg)\\
\nonumber &\hspace{0.5in}
+\left(\prod_{r=t+1}^T \Pr^{(j)}(Z_{j,r}=0)\right)
\Pr\left(X_{j,t:T}|Z_{j,t}=1,Z_{j,(t+1):T}=0\right)\\
&=\left(\sum_{s=t+1}^T \left(\prod_{r=t+1}^{s-1} (1-c_j(r))\right)c_j(s)
P_j(t,s)Q_j(s)\right)+\left(\prod_{r=t+1}^T (1-c_j(r))\right)
P_j(t,T+1).\label{eqn:Qt}
\end{align}
\eqnref{Qt} holds also for $t=1$, by the same derivation.
\eqnsref{QT} and
\eqnssref{Qt} allow us to compute $Q_j(t)$ for $t=T,T-1,T-2,\ldots,1$ recursively
via a ``backward pass''. We may then sample each successive location where
$Z_{j,t}=1$,
conditional on the data $X$ and $Z_{(-j),\cdot}$, in a ``forward pass'':
\begin{align}
\nonumber &\Pr^{(j)}\left(Z_{j,1:(t-1)}=0,Z_{j,t}=1|X\right)\\
\nonumber
&\hspace{0.5in}=\Pr^{(j)}\left(Z_{j,1:(t-1)}=0,Z_{j,t}=1|X_{j,1:T}\right)\\
\nonumber &\hspace{0.5in}=
\tfrac{\Pr^{(j)}(X_{j,1:T}|Z_{j,1:(t-1)}=0,Z_{j,t}=1)
\Pr^{(j)}(Z_{j,1:(t-1)}=0,Z_{j,t}=1)}{\Pr^{(j)}(X_{j,1:T})}\\
\nonumber &\hspace{0.5in}=
\tfrac{\Pr(X_{j,1:(t-1)}|Z_{j,1:(t-1)}=0,Z_{j,t}=1)
\Pr^{(j)}(X_{j,t:T}|Z_{j,t}=1)\left(\prod_{r=2}^{t-1} \Pr^{(j)} Z_{j,r}=0\right)
\Pr^{(j)}(Z_{j,t}=1)}{\Pr^{(j)}(X_{j,1:T})}\\
\label{eqn:rowprob1} &\hspace{0.5in}=
\tfrac{P_j(1,t)Q_j(t)\left(\prod_{r=2}^{t-1} (1-c_j(r))\right)c_j(t)}{Q_j(1)},
\\
\nonumber &\Pr^{(j)}(Z_{j,(s+1):(t-1)}=0,Z_{j,t}=1|Z_{j,s}=1,X,Z_{j,1:(s-1)})\\
\nonumber &\hspace{0.5in}=
\Pr^{(j)}(Z_{j,(s+1):(t-1)}=0,Z_{j,t}=1|Z_{j,s}=1,X_{j,s:T})\\
\nonumber &\hspace{0.5in}=
\tfrac{\Pr^{(j)}\left(X_{j,s:T}|Z_{j,s}=1,Z_{j,(s+1):(t-1)}=0,Z_{j,t}=1\right)
\Pr^{(j)}(Z_{j,(s+1):(t-1)}=0,Z_{j,t}=1|Z_{j,s}=1)}
{\Pr^{(j)}\left(X_{j,s:T}|Z_{j,s}=1\right)}\\
&\hspace{0.5in}=
\tfrac{P_j(s,t)Q_j(t)\left(\prod_{r=s+1}^{t-1} (1-c_j(r))\right)c_j(t)}{Q_j(s)}.
\label{eqn:rowprob2}
\end{align}
To summarize, the procedure to sample $Z_{j,\cdot}|X,Z_{(-j),\cdot}$ is as
follows:
\begin{enumerate}
\item For each $t=2,\ldots,T$, compute $c_j(t)$ according to \eqnref{ct}.
\item (Backward pass) For each $t=T,\ldots,1$, compute $Q_j(t)$ according to
\eqnsref{QT} and \eqnssref{Qt}.
\item (Forward pass) Sample the smallest $t$ for which $Z_{j,t}=1$ according to
\eqnref{rowprob1}. Sample each subsequent $t$ for which $Z_{j,t}=1$
according to \eqnref{rowprob2}.
\end{enumerate}

Regarding computational cost, let us assume that $P_j(t,s)$ may be updated from
$P_j(t,s-1)$ in constant time, as is true for all of the parametric models in
\eqnsref{normalfixedvar}--\eqnssref{laplace}. Then
computing the value of $c_j(t)$ for $t=2,\ldots,T$ in step (1) above takes
$O(T)$ time. For step (2), the value of the summand for each $s=t+1,\ldots,T$
in \eqnref{Qt} may be updated from that for $s-1$
in constant time, so each $Q_j(t)$ may be computed in $O(T)$ time, and step (2)
may be performed in $O(T^2)$ time. Finally, the value in the numerator of
\eqnsref{rowprob1} and \eqnssref{rowprob2} for each $t=2,\ldots,T$ may be
updated from that for $t-1$ in constant time, so step (3) may be performed in
$O(T)$ time. Hence, sampling $Z_{j,\cdot}|X,Z_{(-j),\cdot}$
for all sequences $j=1,\ldots,J$ may be performed in $O(JT^2)$ time.

We next describe the modification of this sampling algorithm to sample
each row $Z_{j,\cdot}$ in a block-wise fashion, by dividing each row
$Z_{j,\cdot}$ into $K$ blocks
$Z_{j,1:(t_1-1)},Z_{j,t_1:(t_2-1)},\ldots,Z_{j,(t_{K-1}:T)}$ and Gibbs sampling
the blocks sequentially. Let $r_j(k)=\max\{r<t_k:Z_{j,r}=1\}$, and let
$s_j(k)=\min\{s\geq t_{k+1}:Z_{j,s}=1\}$,
with the conventions $r_j(k)=1$ if $Z_{j,1:(t_k-1)}=0$ and $s_j(k)=T+1$ if
$Z_{j,t_{k+1}:T}=0$. Let $\Pr^{(j,k)}$ denote probability conditional on
$Z_{j,1:(t_k-1)}$, $Z_{j,t_{k+1}:T}$, and $Z_{(-j),\cdot}$. (Note then that
$r_j(k)$ and $s_j(k)$ are deterministic under $\Pr^{(j,k)}$.) Let
$Q_{j,k}(t)=\Pr^{(j,k)}(X_{j,t:(s_j(k)-1)}|Z_{j,t}=1)$ for $t_k \leq t \leq
t_{k+1}-1$, and $Q_{j,k}(t_k-1)=\Pr^{(j,k)}(X_{j,r_j(k):(s_j(k)-1)})$. Then, in
the backward pass, we may compute
\begin{align*}
Q_{j,k}(t_{k+1}-1)&=P_j(t_{k+1}-1,s_j(k)),\\
Q_{j,k}(t)&=\left(\sum_{s=t+1}^{t_{k+1}-1}
\left(\prod_{r=t+1}^{s-1}(1-c_j(r))\right)c_j(s)P_j(t,s)Q_{j,k}(s)\right)\\
&\hspace{0.5in}+\prod_{r=t+1}^{t_{k+1}-1}(1-c_j(r))P_j(t,s_j(k))
\text{ for } t_k \leq t<t_{k+1}-1,\\
Q_{j,k}(t_k-1)&=\left(\sum_{s=t_k}^{t_{k+1}-1}
\left(\prod_{r=t_k}^{s-1}(1-c_j(r))\right)c_j(s)P_j(r_j(k),s)Q_{j,k}(s)\right)\\
&\hspace{0.5in}+\prod_{r=t_k}^{t_{k+1}-1}(1-c_j(r))P_j(r_j(k),s_j(k)),
\end{align*}
and sample each successive location where $Z_{j,t}=1$, for
$t \in \{t_k,\ldots,t_{k+1}-1\}$, by
\begin{align*}
&\Pr^{(j,k)}(Z_{j,t_k:(t-1)}=0,Z_{j,t}=1|X)
=\tfrac{P_j(r_j(k),t)Q_{j,k}(t)\left(\prod_{r=t_k}^{t-1} (1-c_j(r))
\right)c_j(t)}{Q_{j,k}(t_k-1)},\\
&\Pr^{(j,k)}(Z_{j,(s+1):(t-1)}=0,Z_{j,t}=1|Z_{j,s}=1,X,Z_{j,t_k:(s-1)})
=\tfrac{P_j(s,t)Q_{j,k}(t)\left(\prod_{r=s+1}^{t-1} (1-c_j(r))
\right)c_j(t)}{Q_{j,k}(s)}.
\end{align*}
The derivations of these expressions are similar to those for
\eqnsref{QT}--\eqnssref{rowprob2}, and we omit them for brevity.

The time required to sample each block of
changepoint variables $Z_{j,t_k:(t_{k+1}-1)}$ is $O((t_{k+1}-t_k)^2)$, reducing
the time required to sample all blocks of $Z_{j,\cdot}$ to $O(T)$ if the block
sizes are $O(1)$. Then the total computational cost of sampling
$Z_{j,\cdot}|X,Z_{(-j),\cdot}$ for all sequences $j=1,\ldots,J$ is reduced from
$O(JT^2)$ to $O(JT)$.\\

\noindent \emph{Step 2: Gibbs sampling by columns}\\

To sample each column $Z_{\cdot,t}$ conditional on the remaining columns
$Z_{\cdot,(-t)}$, let $r_t(j)$ and $s_t(j)$ denote the changepoints in the
$j^\text{th}$ sequence immediately before and after time $t$, i.e.,
$r_t(j)=\max\{r:r<t,Z_{j,r}=1\}$ and $s_t(j)=\min\{s:s>t,Z_{j,s}=1\}$,
with the conventions $r_t(j)=1$ if $Z_{j,1:(t-1)}=0$ and $s_t(j)=T+1$ if
$Z_{j,(t+1):T}=0$. Let $\Pr^{(t)}$ denote probability conditional on
$Z_{\cdot,(-t)}$ with associated conditional expectation $\E^{(t)}$. Note that
$r_t(j)$ and $s_t(j)$ are deterministic under $\Pr^{(t)}$. Let
\begin{align}
A_t(j)&=\Pr^{(t)}(X_{j,r_t(j):(s_t(j)-1)}|Z_{j,t}=1)
=P_j(r_t(j),t)P_j(t,s_t(j)),\label{eqn:Aj}\\
B_t(j)&=\Pr^{(t)}(X_{j,r_t(j):(s_t(j)-1)}|Z_{j,t}=0)=P_j(r_t(j),s_t(j))
\label{eqn:Bj}
\end{align}
for each $j=1,\ldots,J$, where $P_j(t,s)$ is as defined in \eqnref{P}. For
each $j=1,\ldots,J$ and each $k=0,\ldots,J-j$, let $R_t(j,k)$ be the
coefficient of $x^ky^{J-j-k}$ in the polynomial $\prod_{i=j+1}^J
(A_t(i)x+B_t(i)y)$,
with the convention $R_t(J,0)=1$. We may compute all of the $R_t(j,k)$
values recursively for $j=J,J-1,\ldots,1$ in an ``upward pass'':
\begin{align}
R_t(J,0)&=1\label{eqn:RJ}\\
R_t(j,k)=&\begin{cases} B_t(j)R_t(j+1,0) & k=0\\
B_t(j)R_t(j+1,k)+A_t(j)R_t(j+1,k-1) & 1 \leq k \leq J-j-1\\
A_t(j)R_t(j+1,J-j-1) & k=J-j.\end{cases}\label{eqn:Rj}
\end{align}

Let
$N_t(j)=\sum_{i=1}^{j-1} Z_{i,t}$ denote the number of changepoints at position
$t$ in sequences 1 to $j-1$, with $N_t(1)=0$. Then
\begin{align*}
&\Pr^{(t)}(q_t=q|Z_{1:(j-1),t},X_{(j+1):J,\cdot})\\
&\hspace{0.5in}\propto \Pr^{(t)}(X_{(j+1):J,\cdot}|q_t=q,Z_{1:(j-1),t})
\Pr^{(t)}(Z_{1:(j-1),t}|q_t=q)\Pr^{(t)}(q_t=q)\\
&\hspace{0.5in}=\left(\prod_{i=j+1}^J \Pr^{(t)}(X_{i,\cdot}|q_t=q)\right)
\Pr(Z_{1:(j-1),t}|q_t=q)\Pr(q_t=q)\\
&\hspace{0.5in}=\Bigg(\prod_{i=j+1}^J
\left(\Pr^{(t)}(X_{i,\cdot}|Z_{j,t}=1,q_t=q)\Pr^{(t)}(Z_{j,t}=1|q_t=q)\right.\\
&\hspace{1in}\left.
+\Pr^{(t)}(X_{i,\cdot}|Z_{j,t}=0,q_t=q)\Pr^{(t)}(Z_{j,t}=0|q_t=q)\right)\Bigg)
\Pr(Z_{1:(j-1),t}|q_t=q)\Pr(q_t=q)\\
&\hspace{0.5in}\propto \left(\prod_{i=j+1}^J
(A_t(i)q+B_t(i)(1-q))\right)q^{N_t(j)}(1-q)^{j-1-N_t(j)}\Pr(q_t=q).
\end{align*}
Letting $c_t(j)=\Pr^{(t)}(Z_{j,t}=1|Z_{1:(j-1),t},X_{(j+1):J,\cdot})
=\E^{(t)}[q_t|Z_{1:(j-1),t},X_{(j+1):J,\cdot}]$, this implies
\begin{align}
\nonumber c_t(j) &=\frac{\int \left(\prod_{i=j+1}^J
(A_t(i)q+B_t(i)(1-q))\right)
q^{N_t(j)+1}(1-q)^{j-1-N_t(j)}\pi_Q(dq)}{\int
\left(\prod_{i=j+1}^J (A_t(i)q+B_t(i)(1-q))\right)
q^{N_t(j)}(1-q)^{j-1-N_t(j)}\pi_Q(dq)}\\
\nonumber &=\frac{\sum_{k=0}^{J-j} \left(R_t(j,k)\int q^{N_t(j)+k+1}
(1-q)^{J-N_t(j)-k-1}\pi_Q(dq)\right)}
{\sum_{k=0}^{J-j} \left(R_t(j,k)\int q^{N_t(j)+k}
(1-q)^{J-N_t(j)-k-1}\pi_Q(dq)\right)}\\
&=\frac{\sum_{k=0}^{J-j} R_t(j,k)f(N_t(j)+k+1)}
{\sum_{k=0}^{J-j} R_t(j,k)g(N_t(j)+k+1)},\label{eqn:cj}
\end{align}
where $f(\cdot)$ and $g(\cdot)$ are as in \eqnsref{fk}--\eqnssref{gk}.
We may then sequentially sample $Z_{1,t},\ldots,Z_{J,t}$, conditional on the 
data $X$ and $Z_{\cdot,(-t)}$, in a ``downward pass'':
\begin{align}
\nonumber &\Pr^{(t)}(Z_{j,t}=1|Z_{1:(j-1),t},X)\\
\nonumber
&\hspace{0.5in}=\Pr^{(t)}(Z_{j,t}=1|Z_{1:(j-1),t},X_{j,r_t(j):(s_t(j)-1)},
X_{(j+1):J,\cdot})\\
\nonumber &\hspace{0.5in}
=\tfrac{\Pr^{(t)}(X_{j,r_t(j):(s_t(j)-1)}|Z_{j,t}=1,Z_{1:(j-1),t},
X_{(j+1):J,\cdot})
\Pr^{(t)}(Z_{j,t}=1|Z_{1:(j-1),t},X_{(j+1):J,\cdot})}
{\Pr^{(t)}(X_{j,r_t(j):(s_t(j)-1)}|Z_{1:(j-1),t},X_{(j+1):J,\cdot})}\\
&\hspace{0.5in}=\tfrac{A_t(j)c_t(j)}{A_t(j)c_t(j)+B_t(j)(1-c_t(j))}.
\label{eqn:columnprob}
\end{align}
To summarize, the procedure to sample $Z_{\cdot,t}|Z_{\cdot,(-t)}$ is as
follows:
\begin{enumerate}
\item For each $j=1,\ldots,J$, compute $A_t(j)$ and $B_t(j)$ according to
\eqnsref{Aj} and \eqnssref{Bj}.
\item (Upward pass) For each $j=J,\ldots,1$ and $k=0,\ldots,J-j$, compute
$R_t(j,k)$ according to \eqnsref{RJ} and \eqnssref{Rj}.
\item (Downward pass) For each $j=1,\ldots,J$, compute $c_t(j)$ according to
\eqnref{cj}, and sample $Z_{j,t}$ according to \eqnref{columnprob}.
\end{enumerate}

Regarding computational cost, computation of $A_t(j)$ and
$B_t(j)$ for $j=1,\ldots,J$ in step (1) requires $O(J)$ time if we compute
the values of $P_j(r,t)$ and $P_j(t,s)$ by updating them from $P_j(r,t-1)$ and
$P_j(t-1,s)$. In step (2), computation of $R_t(j,k)$ for $j=J,\ldots,1$ and
$k=0,\ldots,J-j$ may be performed in $O(J^2)$ time. In step (3),
computation of $c_t(j)$ for a single value of $j$ may be performed in $O(J)$
time, so step (3) may also be performed in $O(J^2)$ time. Hence, sampling
$Z_{\cdot,t}|X,Z_{\cdot,(-t)}$ for all positions $t=2,\ldots,T$ may be performed
in $O(J^2T)$ time.

A computational shortcut is provided by noting that the
sums in the numerator and denominator of \eqnref{cj} typically decay rapidly
as $k$ increases; 
this is theoretically justified by the fact that for each $t$ and $j$,
$(R_t(j,k))_{k=0}^{J-j}$ is a log-concave sequence (being the coefficients of a 
real polynomial with real roots, see Theorem 2 of \citep{stanley}) and that the
mode of this sequence occurs near $k=0$ if most sequences do not
provide evidence of a changepoint at position $t$. Hence in practice we
truncate these sums in step (3) when the size of the summand falls
below a small threshold, and we compute and store the values $R_t(j,k)$ in
step (2) via lazy evaluation, only as they are needed in step (3). We
observe empirically that this yields a very significant reduction in
computational time and does not affect the results of posterior inference.\\

\noindent \emph{Step 3: Swapping columns by Metropolis-Hastings}\\

Let $P_j(t,s)$ be as defined in \eqnref{P}.
The following describes a Metropolis-Hastings move that potentially swaps two
adjacent columns of the changepoint variable matrix $Z$:

\begin{enumerate}
\item Let $\mathcal{T}=\{t:\sum_{j=1}^J Z_{j,t}>0\}$ be the set of positions
where there is at least one changepoint. Select $t$ uniformly at
random from $\mathcal{T}$, and set $t'=t-1$ or $t'=t+1$ randomly with
probability $\frac{1}{2}$ each. If $t=T$, set $t'=t-1$ with probability 1, and
if $t=2$, set $t'=t+1$ with probability 1. (Recall that in our notation,
$Z_{\cdot,t}=0$ is fixed for $t=1$.)
\item For each $j=1,\ldots,J$, if $Z_{j,t} \neq Z_{j,t'}$, let
$r(j)=\max\{r:r<(t \wedge t'),Z_{j,r}=1\}$, and let
$s(j)=\min\{s:s>(t \vee t'),Z_{j,s}=1\}$, with the conventions $r(j)=1$ if
$Z_{j,1:(t \wedge t')}=0$ and $s(j)=T+1$ if $Z_{j,(t\vee t'):T}=0$.
Compute
\[p\defeq\prod_{j:Z_{j,t}=1,Z_{j,t'}=0}
\frac{P_j(r(j),t')P_j(t',s(j))}{P_j(r(j),t)P_j(t,s(j))}
\prod_{j:Z_{j,t}=0,Z_{j,t'}=1}
\frac{P_j(r(j),t)P_j(t,s(j))}{P_j(r(j),t')P_j(t',s(j))}.\]
\item If $\sum_{j=1}^J Z_{j,t'}>0$, or if
$(t,t') \notin \{(2,3),(3,2),(T-1,T),(T,T-1)\}$, then swap
$Z_{\cdot,t}$ and $Z_{\cdot,t'}$ with probability $\min(p,1)$. If $\sum_{j=1}^J
Z_{j,t'}=0$ and $(t,t') \in \{(2,3),(T,T-1)\}$, then swap $Z_{\cdot,t}$ and
$Z_{\cdot,t'}$ with probability $\min\left(\tfrac{p}{2},1\right)$. Finally, if
$\sum_{j=1}^J Z_{j,t'}=0$ and $(t,t') \in \{(3,2),(T-1,T)\}$, then swap
$Z_{\cdot,t}$ and $Z_{\cdot,t'}$ with probability $\min(2p,1)$.
\end{enumerate}

To see that this procedure keeps the posterior distribution invariant, let
$\tilde{Z}$ denote $Z$ with columns $t$ and $t'$ swapped. Note that under the
BASIC model, $\Pr(Z)=\Pr(\tilde{Z})$. Then the
quantity $p$ computed in step (2) above is precisely
\[p=\frac{\Pr(X|\tilde{Z})}{\Pr(X|Z)}
=\frac{\Pr(X,\tilde{Z})}{\Pr(X,Z)}
=\frac{\Pr(\tilde{Z}|X)}{\Pr(Z|X)}.\]

\noindent The procedure of selecting $(t,t')$ in step (1) induces a transition
probability $Z \to \tilde{Z}$, where $\Pr(Z \to \tilde{Z})
=\Pr(\tilde{Z} \to Z)$ in most cases, with the exceptions
$\Pr(Z \to \tilde{Z})=\frac{1}{|\mathcal{T}|}$
and $\Pr(\tilde{Z} \to Z)=\frac{1}{2|\mathcal{T}|}$
if $\sum_{j=1}^J Z_{j,t'}=0$ and $(t,t')=(2,3)$ or $(T,T-1)$, and
$\Pr(Z \to \tilde{Z})=\frac{1}{2|\mathcal{T}|}$ and $\Pr(\tilde{Z} \to
Z)=\frac{1}{|\mathcal{T}|}$ if $\sum_{j=1}^J Z_{j,t'}=0$ and $(t,t')=(3,2)$
or $(T-1,T)$. Step (3) above handles all cases with the correct
Metropolis-Hastings acceptance probability. In practice, the most common
scenario is when there are no changepoints at position $t'$, in which case the
``swap'' of columns $t$ and $t'$ simply shifts all changepoints at position $t$
by one position.

Regarding computational cost, to perform the above
procedure, one may precompute $P_j(t,s)$ for each sequence $j$ and each pair of
consecutive changepoints $t,s$ in sequence $j$ (i.e., $Z_{j,t}=1$,
$Z_{j,(t+1):(s-1)}=0$, and $Z_{j,s}=1$). This requires $O(JT)$ computational
cost. Then step (1) above requires $O(1)$ cost, step (2) requires $O(J)$ cost,
and step (3) requires $O(J)$ cost. Upon performing the swap in step (3), the set
$\mathcal{T}$ and the values $P_j(t,s)$ may easily be updated in $O(J)$ time, to
prepare for the next application of this Metropolis-Hastings move. Hence,
performing $B$ total iterations of the Metropolis-Hastings move requires
$O(JT+JB)$ time. In our applications we set $B=10T$, and we observe that the
computational
cost of performing all $B$ Metropolis-Hastings steps is much smaller than the
cost of the row-wise and column-wise Gibbs sampling procedures.

\section{Posterior maximization
algorithms}\label{sec:maximizationalgorithms}
Below are the details of the iterative posterior maximization algorithm
discussed in Section \ref{subsecmaximization}.\\

\noindent \emph{Step 1: Maximizing over rows}\\

Note that $\Pr(Z|X)=\Pr(Z_{j,\cdot}|X,Z_{(-j),\cdot})\Pr(Z_{(-j),\cdot}|X)$,
so maximizing $\Pr(Z|X)$ over the row $Z_{j,\cdot}$ is equivalent to
maximizing $\Pr(Z_{j,\cdot}|X,Z_{(-j),\cdot})$. To perform this maximization, we
may employ the dynamic programming recursions developed by Brad Jackson et al.\ 
for the univariate changepoint problem \citep{jacksonetal}, in the following
way.

Note that
\begin{align}
\nonumber \Pr(Z_{j,\cdot}|X,Z_{(-j),\cdot})
&=\Pr(Z_{j,\cdot}|X_{j,\cdot},Z_{(-j),\cdot})\\
\nonumber &\propto \Pr(X_{j,\cdot}|Z_{j,\cdot})\Pr(Z_{j,\cdot}|Z_{(-j),\cdot})\\
\nonumber &=\Pr(X_{j,\cdot}|Z_{j,\cdot})\prod_{t=2}^T
\left(\Pr[Z_{j,t}=1|Z_{(-j),\cdot}]^{Z_{j,t}}
(1-\Pr[Z_{j,t}=1|Z_{(-j),\cdot}])^{1-Z_{j,t}}\right)\\
&=\Pr(X_{j,\cdot}|Z_{j,\cdot})\prod_{t=2}^T
c_j(t)^{Z_{j,t}}(1-c_j(t))^{1-Z_{j,t}},
\label{eqn:rowcondprob}
\end{align}
where $c_j(t)=\Pr[Z_{j,t}=1|Z_{(-j),\cdot}]$ may be computed as \eqnref{ct}.
Define $M_j(1)=\Pr(X_{j,1}|Z_{j,2}=1)$, the marginal probability
density of the first data point in sequence $j$ assuming there is a
changepoint immediately after it, and for $t=2,\ldots,T$, define
\begin{align*}
V_{j,t}(Z_{j,1:t})&=\Pr(X_{j,1:t}|Z_{j,1:t},Z_{j,t+1}=1)\prod_{r=2}^t
c_j(r)^{Z_{j,r}}(1-c_j(r))^{1-Z_{j,r}},\\
M_j(t)&=\max_{Z_{j,1:t}} V_{j,t}(Z_{j,1:t}).
\end{align*}
Then \eqnref{rowcondprob} is exactly $V_{j,T}(Z_{j,1:T})$, and we wish to
compute the sequence $Z_{j,1:T}$ that achieves the maximal value $M_j(T)$.
We do this by iteratively computing $M_j(t)$ for $t=1,\ldots,T$.

Let $R_j(t,1)=V_{j,t}((0,0,\ldots,0))$ be the value of $V_{j,t}$ if there are no
changepoints before position $t$ in sequence $j$, and for $s=2,\ldots,t$, let
\[R_j(t,s)=\max_{Z_{j,1:t}:\;Z_{j,s}=1,Z_{j,(s+1):t}=0}
V_{j,t}(Z_{j,1:t})\]
be the maximal value of $V_{j,t}$ assuming that the last changepoint in sequence
$j$ before position $t$ occurs at position $s$. Then, with $P_j(t,s)$ as in
\eqnref{P},
\begin{align}
M_j(1)&=P_j(1,2),\label{eqn:M1}\\
R_j(t,1)&=P_j(1,t+1)\prod_{r=2}^t (1-c_j(r)),\label{eqn:Rt1}\\
\nonumber R_j(t,s)&=\max_{Z_{j,1:(s-1)}} \left(\Pr(X_{j,1:(s-1)}|Z_{j,1:(s-1)},
Z_{j,s}=1)\prod_{r=2}^{s-1}c_j(r)^{Z_{j,r}}(1-c_j(r))^{1-Z_{j,r}}\right)\times\\
\nonumber &\hspace{0.5in}\Pr(X_{j,s:t}|Z_{j,s}=1,Z_{j,(s+1):t}=0,Z_{j,t+1}=1)
c_j(s)\prod_{r=s+1}^t(1-c_j(r))\\
&=M_j(s-1)P_j(s,t+1)c_j(s)\prod_{r=s+1}^t (1-c_j(r)),\label{eqn:Rts}\\
M_j(t)&=\max_{s=1,\ldots,t} R_j(t,s).\label{eqn:Mt}
\end{align}
The above recursions are similar to those in Section II of \citep{jacksonetal}.
From these recursions, we may compute $M_j(t)$ for each $t=2,\ldots,T$ by
computing $R_j(t,s)$ for each $s=1,\ldots,t$. In the sequence $Z_{j,1:T}$ that 
achieves the maximum value $M_j(T)$,
the last changepoint is the index $t$ such that
$M_j(T)=R_j(T,t)$, the changepoint before $t$ is the index $s$ such that
$M_j(t-1)=R_j(t-1,s)$, etc.

To summarize, the procedure to maximize $\Pr(Z_{j,\cdot}|X,Z_{(-j),\cdot})$
over $Z_{j,\cdot}$ is as follows:

\begin{enumerate}
\item For each $t=2,\ldots,T$, compute $c_j(t)$ according to \eqnref{ct}.
\item Compute $M_j(1)$ according to \eqnref{M1}. For each
$t=2,\ldots,T$, compute $R_j(t,s)$ for $s=1,\ldots,t$ according
to \eqnsref{Rt1} and \eqnssref{Rts}, and then compute $M_j(t)$ according
to \eqnref{Mt}. For each $t$, save the value of $s$ such that
$M_j(t)=R_j(t,s)$.
\item Let $\mathcal{S}=\{T+1\}$. While the smallest value in $\mathcal{S}$ is
greater than 1, let this smallest value be $t$, let $s$ be the value that
achieved $M_j(t-1)=R_j(t-1,s)$, update $\mathcal{S} \to \mathcal{S} \cup \{s\}$,
and repeat. When the smallest value in $\mathcal{S}$ becomes 1, set
$Z_{j,t}=1$ for each $t \in \mathcal{S}$ with $2 \leq t \leq T$, and set
$Z_{j,t}=0$ for all other $t$.
\end{enumerate}

Regarding the computational cost,
computation of $c_j(t)$ for $t=2,\ldots,T$ in step (1) above requires
$O(T)$ time. For step (2), $R_j(t,1)$ may be computed in $O(T)$
time for each $t$, and $R_j(t,s)$ may be updated from $R_j(t,s-1)$ in constant
time for each $s=2,\ldots,t$, so all of the values $R_j(t,s)$ and $M_j(t)$ for
$t=2,\ldots,T$ and $s=1,\ldots,t$ in step (2) may be computed in $O(T^2)$ time.
Since step (3) may be performed in $O(T)$ time, maximizing
$\Pr(Z_{j,\cdot}|X,Z_{(-j),\cdot})$ over $Z_{j,\cdot}$ for all $j=1,\ldots,J$
may be performed in $O(JT^2)$ time.

We next describe the modification of this maximization algorithm to maximize
over each row $Z_{j,\cdot}$ in a block-wise fashion, by dividing each row
$Z_{j,\cdot}$ into $K$ blocks
$Z_{j,1:(t_1-1)},Z_{j,t_1:(t_2-1)},\ldots,Z_{j,(t_{K-1}:T)}$ and maximizing over
the blocks sequentially. Let $r_j(k)=\max\{r<t_k:Z_{j,r}=1\}$,
and let $s_j(k)=\min\{s\geq t_{k+1}:Z_{j,s}=1\}$, with the conventions
$r_j(k)=1$ if $Z_{j,1:(t_k-1)}=0$ and $s_j(k)=T+1$ if $Z_{j,t_{k+1}:T}=0$.
Then we may set $M_{j,k}(t_k-1)=P_j(r_j(k),t_k)$ and compute recursively for
$t=t_k,\ldots,t_{k+1}-1$ and $s=t_k,\ldots,t$
\begin{align*}
R_{j,k}(t,t_k-1)&=\begin{cases}
P_j(r_j(k),t+1)\prod_{r=t_k}^t (1-c_j(r)) & t=t_k,\ldots,t_{k+1}-2\\
P_j(r_j(k),s_j(k))\prod_{r=t_k}^{t_{k+1}-1} (1-c_j(r)) & t=t_{k+1}-1,
\end{cases}\\
R_{j,k}(t,s)&=\begin{cases}
M_{j,k}(s-1)P_j(s,t+1)c_j(s)\prod_{r=s+1}^t (1-c_j(r))
& t=t_k,\ldots,t_{k+1}-2\\
M_{j,k}(s-1)P_j(s,s_j(k))c_j(s)\prod_{r=s+1}^{t_{k+1}-1}
(1-c_j(r)) & t=t_{k+1}-1,
\end{cases}\\
M_{j,k}(t)&=\max_{s=t_k-1,\ldots,t} R_j(t,s).
\end{align*}
The interpretations and derivations of the above expressions are similar to
those for \eqnsref{rowcondprob}--\eqnssref{Mt}, and we omit them for
brevity. Then, initializing $\mathcal{S}=\{t_{k+1}\}$, we may iteratively take
the smallest value $t$ in $\mathcal{S}$, let $s$ be such that
$M_{j,k}(t-1)=R_{j,k}(t-1,s)$, update $\mathcal{S} \to \mathcal{S} \cup \{s\}$,
and repeat until $s=t_k-1$, to obtain $Z_{j,t_k:(t_{k+1}-1)}$ that
maximizes the posterior probability over this block.

The time required to maximize over each
block $Z_{j,t_k:(t_{k+1}-1)}$ is $O((t_{k+1}-t_k)^2)$,
reducing the time required to maximize over all blocks of $Z_{j,\cdot}$ to
$O(T)$ if the block sizes are $O(1)$. Then the total computational cost of
maximizing over $Z_{j,\cdot}$ for all sequences $j=1,\ldots,J$ is reduced
from $O(JT^2)$ to $O(JT)$.\\

\noindent \emph{Step 2: Maximizing over columns}\\

Note that $\Pr(Z|X)=\Pr(Z_{\cdot,t}|X,Z_{\cdot,(-t)})\Pr(Z_{\cdot,(-t)}|X)$,
so maximizing $\Pr(Z|X)$ over the column $Z_{\cdot,t}$ is equivalent to
maximizing $\Pr(Z_{\cdot,t}|X,Z_{\cdot,(-t)})$. To perform this maximization,
let $N_t=\sum_{j=1}^J Z_{j,t}$ denote the number of changepoints at position
$t$. Note that $N_t$ is a function of $Z_{\cdot,t}$. Let $r_t(j)$
and $s_t(j)$ denote the changepoints in the $j^\text{th}$ sequence immediately
before and after position $t$, i.e., $r_t(j)=\max\{r:r<t,Z_{j,r}=1\}$ and
$s_t(j)=\min\{s:s>t,Z_{j,s}=1\}$, with the conventions $r_t(j)=1$ if
$Z_{j,1:(t-1)}=0$ and $s_t(j)=T+1$ if $Z_{j,(t+1):T}=0$. Recall the quantities
$A_t(j)$ and $B_t(j)$ from \eqnsref{Aj} and \eqnssref{Bj}. Then
\begin{align*}
\Pr\left(Z_{\cdot,t}|X,Z_{\cdot,(-t)}\right) &\propto
\Pr(X|Z)\Pr\left(Z_{\cdot,t}|Z_{\cdot,(-t)}\right)\\
&\propto \left(\prod_{j:Z_{j,t}=1} A_t(j)\right)
\left(\prod_{j:Z_{j,t}=0} B_t(j)\right)
\sum_{q \in S} \Pr\left(Z_{\cdot,t}|q_t=q\right)\Pr(q_t=q)\\
&\propto \left(\prod_{j:Z_{j,t}=1} \frac{A_t(j)}{B_t(j)}\right)f(N_t),
\end{align*}
where $f(k)$ is defined in \eqnref{fk}.
For any fixed $N_t$, the above quantity is maximized by setting $Z_{j,t}=1$ for
the $N_t$ indices $j \in \{1,\ldots,J\}$ that correspond to the $N_t$ largest
values of $\frac{A_t(j)}{B_t(j)}$, and setting $Z_{j,t}=0$ for all other $j$.
Hence, to maximize $\Pr(Z_{\cdot,t}|X,Z_{\cdot,(-t)})$ over $Z_{\cdot,t}$, we
may perform the following procedure:
\begin{enumerate}
\item For each $j=1,\ldots,J$, compute $\frac{A_t(j)}{B_t(j)}$ according to 
\eqnsref{Aj} and \eqnssref{Bj}, and sort these values.
\item For each $k=0,\ldots,J$, compute the maximum value of
$\left(\prod_{j:Z_{j,t}=1} \frac{A_t(j)}{B_t(j)}\right)f(k)$
over $Z_{\cdot,t}$ such that $\sum_{j=1}^J Z_{j,t}=k$. Let $k^*$ be the value
of $k$ that maximizes this value.
\item Set $Z_{j,t}=1$ for the $k^*$ values of $j$ corresponding to the $k^*$
largest values of $\frac{A_t(j)}{B_t(j)}$, and set $Z_{j,t}=0$ for all other
$j$.
\end{enumerate}

Regarding computation cost, 
$\frac{A_j(t)}{B_j(t)}$ may be computed for $j=1,\ldots,J$ in step (1) in $O(J)$
time, if $A_t(j)$ and $B_t(j)$ are updated from $A_{t-1}(j)$ and $B_{t-1}(j)$,
and they may be sorted in $O(J \log J)$ time. Step (2)
may be performed in $O(J)$ time. Since step (3) also may be performed in
$O(J)$ time, maximizing $\Pr(Z_{\cdot,t}|X,Z_{\cdot,(-t)})$ over
$Z_{\cdot,t}$ for all $t=2,\ldots,T$ may be performed in $O(JT \log J)$ time.\\

\noindent \emph{Step 3: Swapping columns}\\

The following procedure allows for adjustment of all changepoints at a position
$t$ to a new position $t+1$ or $t-1$:
Let $\mathcal{T}=\{t:\sum_{j=1}^J Z_{j,t}>0\}$ be the set of positions
where there is at least one changepoint. For $t \in \mathcal{T}$,
let $Z_+$ denote $Z$ with columns $t$ and $t+1$ swapped, and let
$Z_-$ denote $Z$ with columns $t$ and $t-1$ swapped.
While there exists $t \in \mathcal{T}$ such that
$\Pr(X|Z)$ is less than $\Pr(X|Z_+)$ or $\Pr(X|Z_-)$, update $Z$ to $Z_+$ or
$Z_-$ accordingly, and repeat. Note that as $\Pr(Z|X) \propto \Pr(X|Z)\Pr(Z)$
and $\Pr(Z_+)=\Pr(Z_-)=\Pr(Z)$,
the posterior probability $\Pr(Z|X)$ always increases with each swap. As in the
case of our Metropolis-Hastings move in
Section \ref{subsecsampling}, the primary purpose of this routine is to swap
column $t$ for column $t'=t+1$ or $t'=t-1$ when
$\sum_{j=1}^J Z_{j,t'}=0$, in which case the ``swap'' simply moves all
changepoints at position $t$ to $t'$.

Regarding computational cost, one may precompute $P_j(t,s)$ for each sequence
$j$ and each pair of consecutive changepoints $t,s$ in sequence $j$. This
requires $O(J|\mathcal{T}|)$ computational time where $|\mathcal{T}| \leq T$
is the total number of positions with a changepoint in $Z$.
Then it is evident that
$\frac{\Pr(X|Z_+)}{\Pr(X|Z)}$ and $\frac{\Pr(X|Z_-)}{\Pr(X|Z)}$ may be
computed in $O(J)$ time from these quantities.
Upon performing a swap of, say, $t$ with $t+1$, the new values $P_j(t+1,s)$
and $P_j(s,t+1)$ for changepoints $s$ immediately preceding and following
$t+1$ may be computed in $O(J)$ time, to prepare for evaluation of the next
swap. Hence each swap throughout the procedure
may be performed in $O(J)$ time. In practice, we observe that very few
swaps are made, and the total computational cost of column-swapping is
dominated by the $O(J|\mathcal{T}|)$ initialization time and is also
negligible compared to the costs of row-wise and column-wise maximization
over $Z$.

\section{MCEM algorithms}\label{sec:MCEMalgorithms}
We describe details of the maximization steps in our MCEM procedure.
Maximization over $\eta$ is dependent on the choices of the likelihood model
$p(x|\theta)$ and the prior model $p(\theta|\eta)$. In all of the examples of
\eqnsref{normalfixedvar}--\eqnssref{laplace}, $\eta$ is a low-dimensional
parameter, and a closed-form expression is available for computing
$\log P_j(t,s|\eta)$. We use the BOBYQA zeroth-order
optimization routine \citep{powell}, as implemented in the C++ dlib library, to
maximize over $\eta$.

For the maximization over the probability weights $\{w_k\}_{k \in S}$,
observe that the objective function is a convex function of these weights. In 
fact, define a probability measure $\mu_{\pi_Q}$ on $\{0,\ldots,J\}$ by
\[\mu_{\pi_Q}(j)=\sum_{k \in S} w_k\int \binom{J}{j}q^j(1-q)^{J-j}\nu_k(dq),\]
i.e.\ $\mu_{\pi_Q}(j)$ is the probability under $\pi_Q$ of observing exactly $j$
changepoints at any position $t$. Denote by $\bar{\mu}$ the distribution over
$\{0,\ldots,J\}$ with mass function $\bar{\mu}(j)=\sum_{m=1}^M \frac{N_j^{(m)}}
{M(T-1)}$. (Note that $\sum_{j=0}^J N_j=T-1$ by definition of $N_j$, so
$\sum_{j=0}^J \bar{\mu}(j)=1$.) Then the cross entropy between $\bar{\mu}$ and
$\mu_{\pi_Q}$ is given by
\[-\sum_{j=0}^J \bar{\mu}(j)\log \mu_{\pi_Q}(j)=
-\sum_{j=0}^J \sum_{m=1}^M \frac{N_j^{(m)}}{M(T-1)}\log\left(\sum_{k \in S}
w_k\int \binom{J}{j}q^j(1-q)^{J-j}\nu_k(dq)\right).\]
As this cross entropy is equal to $D_{KL}(\bar{\mu}||\mu_{\pi_Q})+H(\bar{\mu})$,
where $D_{KL}(\bar{\mu}||\mu_{\pi_Q})$ denotes the Kullback-Leibler
divergence and $H(\bar{\mu})$ denotes the Shannon entropy, this implies
\[\frac{1}{M(T-1)}\sum_{m=1}^M \sum_{j=0}^J N_j^{(m)}
\log\left(\sum_{k \in S} w_k \int q^j(1-q)^{J-j}\nu_k(dq)\right)
=-D_{KL}(\bar{\mu}||\mu_{\pi_Q})+\text{const.}\]
for a constant independent of $\pi_Q$. Hence the optimization over $\pi_Q$ may
be written as
\begin{equation}
\{w^{(i)}_k\}_{k \in S}=\argmin_{\{w_k\}}
D_{KL}(\bar{\mu}||\mu_{\pi_Q}).\label{eqn:piqoptKL}
\end{equation}
This may be solved efficiently via an iterative divergence minimization
procedure
\begin{equation}
w^{(i)}_k \leftarrow w^{(i-1)}_k \sum_{j=0}^J
\frac{\bar{\mu}(j)\int q^j(1-q)^{J-j}\nu_k(dq)}{\sum_{k' \in S}
w^{(i-1)}_{k'}\int q^j(1-q)^{J-j}\nu_{k'}(dq)},\label{eqn:piqiteration}
\end{equation}
which converges to the global optimum
in \eqnref{piqoptKL}, provided that it is initialized to a probability
vector supported on all of $S$ \citep{csiszarshields,lashkarigolland}.
To iteratively compute the update in \eqnref{piqiteration}, one may precompute
$\int q_j(1-q)^{J-j}\nu_k(dq)$ for each $j$ and $k$.

In our applications, we take
$\{\nu_k\}_{k \in S}=\{k/J\}_{k=0}^{\lfloor J/2 \rfloor-1}$, and we
initialize $\{w_k^{(0)}\}$ such that $w_0^{(0)}=0.9$ and the remaining
probability mass of 0.1 is spread equally over the other grid points $k/J$.
We initialize $\eta^{(0)}$ by
dividing the data in each sequence into blocks of 100 data points, computing the
sample mean and/or variance within each block, and matching the empirical
moments of these sample means and/or variances to their theoretical moments
under the prior $\pi_\Theta$. For instance, for the
normal model with changing mean, \eqnref{normalfixedvar}, we initialize
$\mu_0$ to the empirical average of the block means, $\sigma_0^2$ to the
empirical average of the block variances, and $\lambda$ to $\sigma_0^2$
divided by the empirical variance of the block means. A similar procedure
is used for the other parametric models of
\eqnsref{normalfixedmean}--\eqnssref{laplace}.

\section{Gibbs sampling comparisons}\label{sec:gibbscomparison}
\begin{figure}[tbp]
\includegraphics[width=\textwidth]{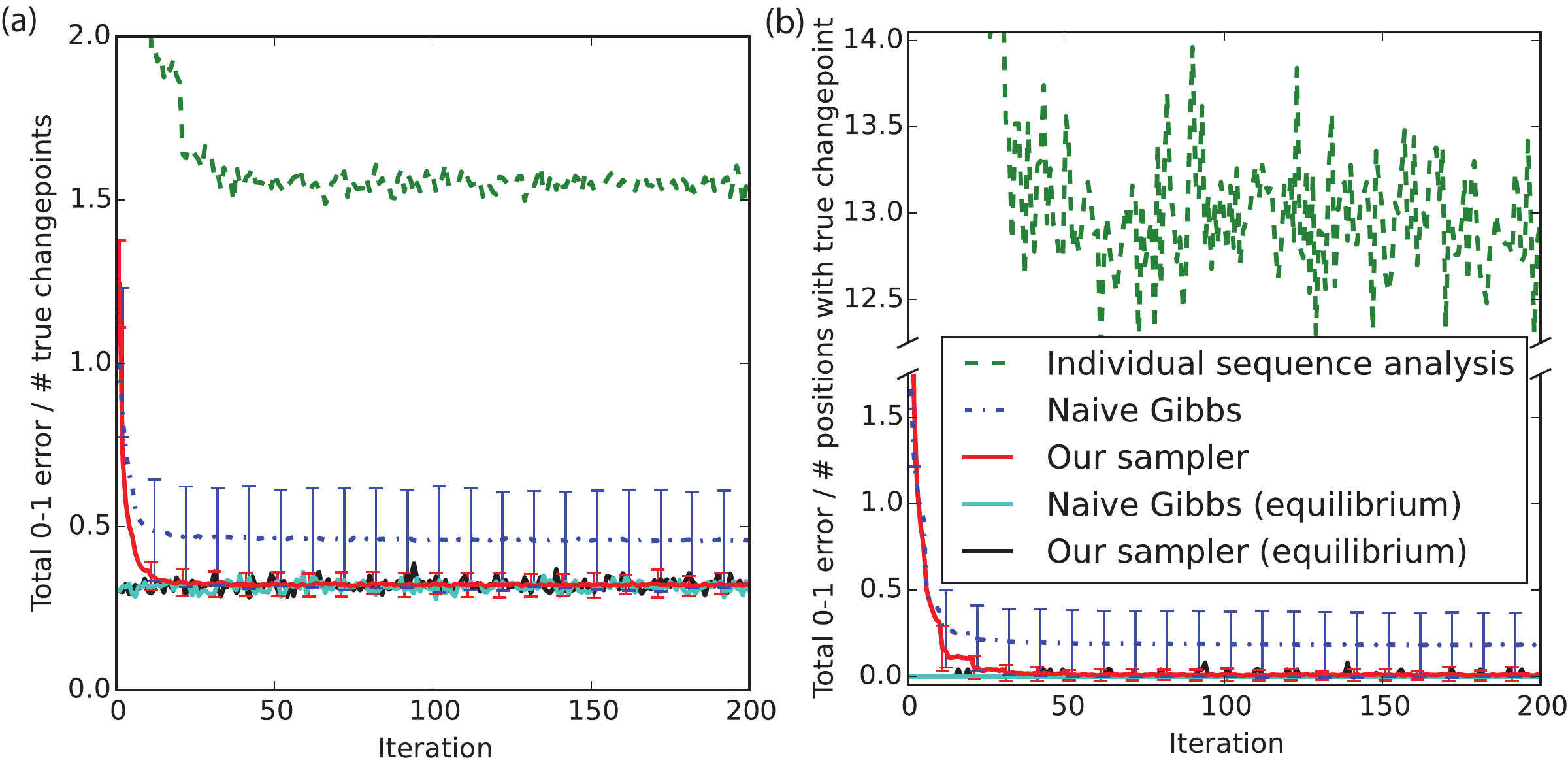}
\caption{
Relative changepoint error (a) and change position error (b) of alternative MCMC inference procedures applied to data generated from the BASIC model.
Also plotted is the aggregated error from one
run of an analysis of each sequence individually.}
\label{fig:gibbscomparison}
\end{figure}
We examine convergence to equilibrium of our MCMC sampling algorithm
on a data set with $J=50$ sequences and $T=10000$ observations per
sequence. We compare the performance of our algorithm with a naive Gibbs
sampler and investigate also the effect of row block size in the accelerated
version of our sampler.
The data was generated according to the BASIC model with true
changepoint prior $\pi_Q=0.995\delta_0+0.005\delta_{0.4}$, using the likelihood
of \eqnref{normalfixedvar} with $\mu_0=0$, $\lambda=1$, and $\sigma_0^2=1$.
The generated data contained 1018
total changepoints at 50 distinct sequential positions.

We performed experiments in which we ran 200 iterations of
the MCMC sampling procedure of Section \ref{subsecsampling}. Prior parameters
were initialized to default settings as discussed in
\secref{MCEMalgorithms} and updated with MCEM after sampling
iterations 5, 10, 20, 30, and 50.
Red lines in Figure \ref{fig:gibbscomparison} depict the error of the sampled
changepoints at each iteration, averaged across 50 independent replicates of
this experiment, with error bars depicting $\pm 2$ standard deviations.
Panel (a) displays the \emph{relative changepoint error},
which is the total 0--1 error of changepoint detections,
divided by 1018 (the total number of true
changepoints). Panel (b) displays the \emph{relative change position error},
which is the 0--1 error of detected sequential positions having a changepoint in
any sequence, divided by 50 (the total number of true sequential positions
having such a change). As a comparison,
the dashed green curve in Figure \ref{fig:gibbscomparison} shows the
errors when each sequence is treated individually as its own data set and
indicates the accuracy of an analogous analysis that does not pool information
across sequences.

Dashed blue curves and error bars in Figure \ref{fig:gibbscomparison}
correspond to the results
of applying a naive Gibbs sampling algorithm to sample from the posterior
distribution under the BASIC model. In this naive sampler, the latent
variables $q_t$ and $\theta_{j,t}$ are still marginalized out analytically,
but the latent changepoint variables $Z_{j,t}$ are individually Gibbs-sampled.
This sampling scheme is easy to implement and does not require the dynamic
programming recursions detailed in \secref{samplingalgorithms}. To equate
runtime with that of our MCMC procedure, 30 iterations of naive Gibbs
sampling are treated as ``one iteration'' in Figure \ref{fig:gibbscomparison}.
We observe that even though many iterations of naive Gibbs sampling can be
performed in the
same amount of time as one iteration of our procedure, the naive
Gibbs sampler did not consistently converge to the same level of error. 

Black and cyan curves in Figure \ref{fig:gibbscomparison} show errors from
a single experiment of our MCMC sampler and the naive Gibbs sampler,
respectively, initialized to the true changepoint
matrix $Z^\true$ and using the true priors $\pi_Q$ and $\pi_\Theta$.
Both curves remain stable around the same
``equilibrium'' error value across all 200 iterations, providing
evidence that the our sampler without this ideal initialization (red
curve) indeed reaches equilibrium sampling of the posterior distribution after
few iterations.

\begin{figure}[tbp]
\includegraphics[width=0.5\textwidth]{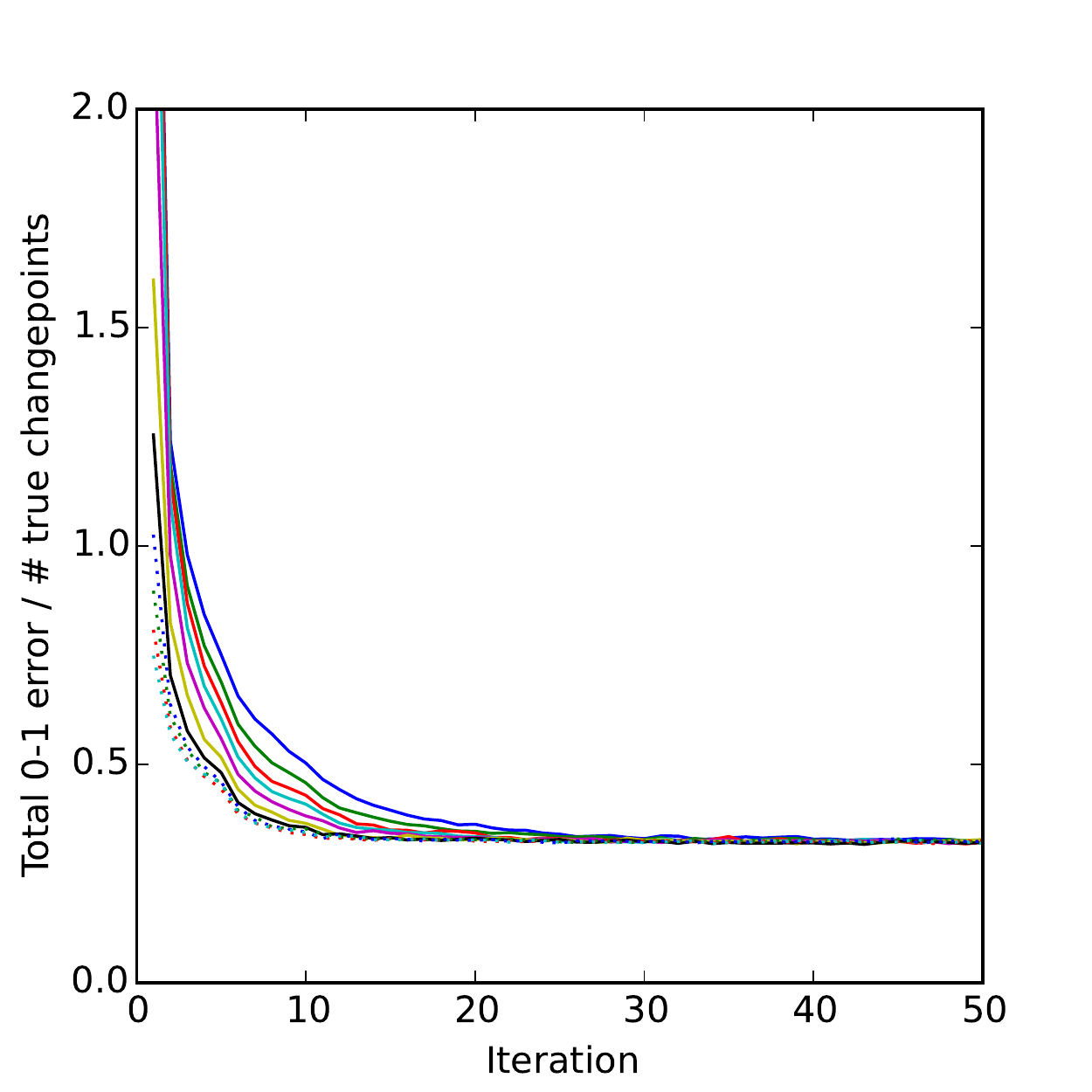}%
\includegraphics[width=0.5\textwidth]{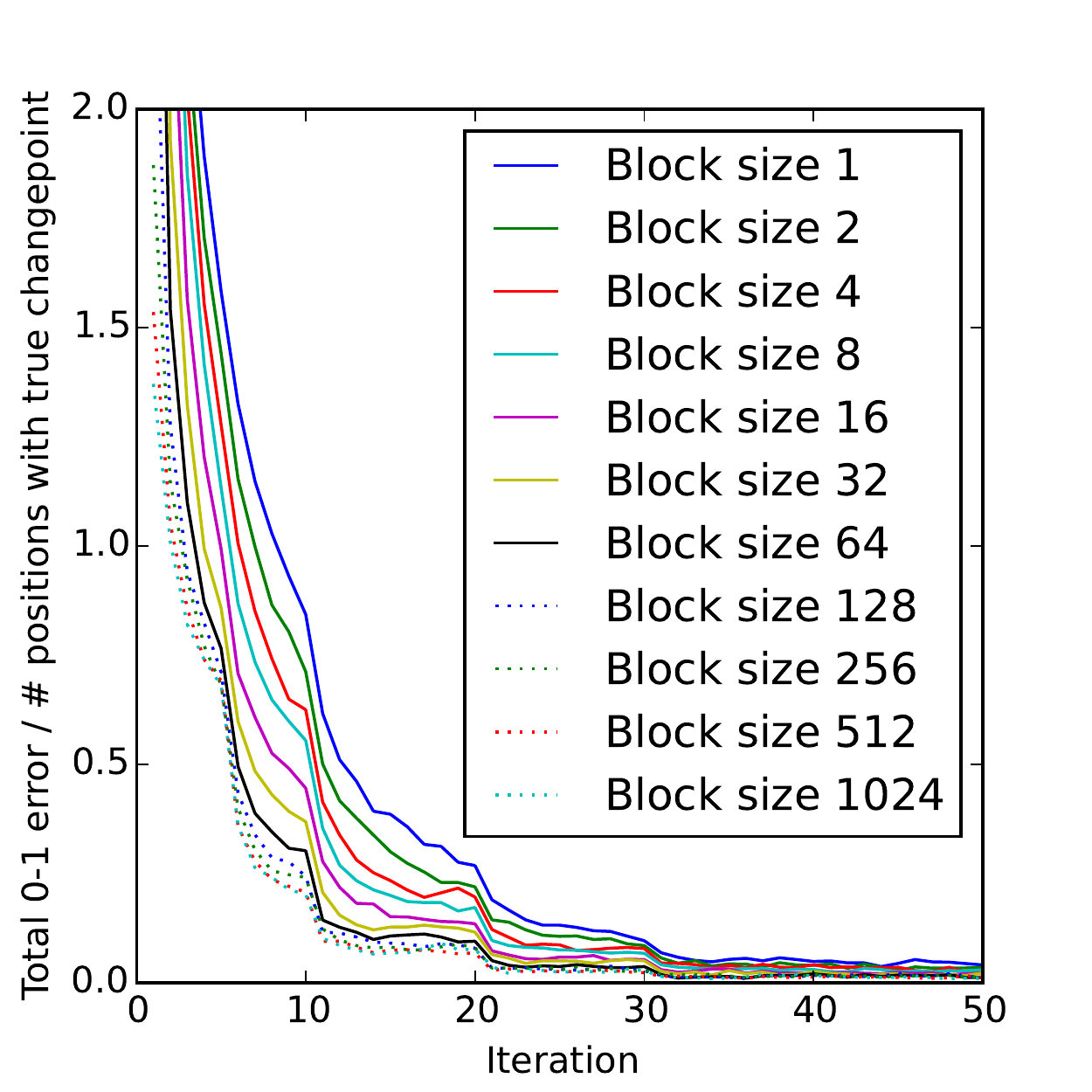}
\caption{Effects of row block size choice on sampling.
Relative changepoint error and change position error are as in Figure
\ref{fig:gibbscomparison}.}
\label{figblocksize}
\end{figure}

In the above comparisons, our MCMC sampler was run with the default setting of
row block size 50 in the acceleration described in Section \ref{seclineartime}.
Figure \ref{figblocksize} explores the effect of this block
size choice on sampling: We tested block sizes in powers of two between 1 and
1024, and the curves correspond to the mean error across 50 independent
experiments for the same two error metrics.
(The sampler with block size 1 is different from the naive Gibbs
sampler above, as we still apply the column-wise Gibbs sampling and
Metropolis-Hastings column swap steps of our procedure.) In this example,
the average spacing between changepoints is 200 across all sequences and 500 in
any particular sequence. We observe that there
is only a small improvement in sampling if block sizes are increased beyond 64;
however, there is a large increase in computational time per iteration. On the
other hand, reducing the block size to be very small does not yield a
substantial reduction in computational time, if the column-wise sampling step is
still applied in each iteration. We believe our default choice of block size 50
is a reasonable setting in most applications.

\section{Comparison of methods on data of \cite{louhimoetal} without
subsampling}\label{sec:syntheticCNV}
\begin{figure}[t]
\begin{minipage}{0.6\textwidth}
\includegraphics[width=\textwidth]{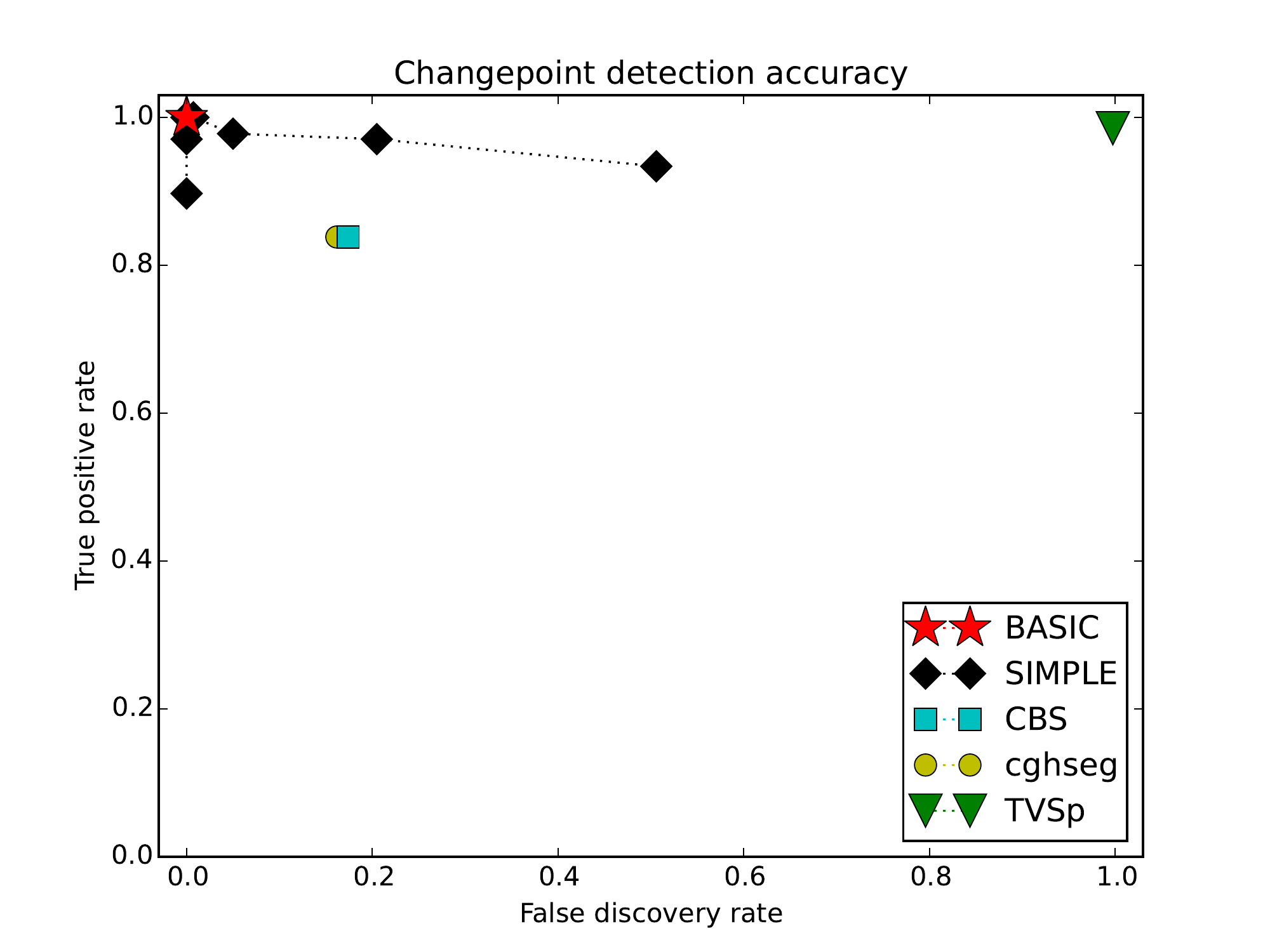}
\end{minipage}%
\begin{minipage}{0.4\textwidth}
\begin{center}
Signal reconstruction error\\
\vspace{0.1in}
\begin{tabular}{c|c}
Method & $\sum_{j,t} (\mu_{j,t}^\text{est}-\mu_{j,t}^\text{true})^2$\\
\hline
BASIC & 7.40 \\
SIMPLE & 7.40 \\
CBS & 16.25 \\
cghseg & 16.11 \\
TVSp & 257.2
\end{tabular}
\end{center}
\end{minipage}
\caption{Changepoint detection accuracy and signal reconstruction squared-error
for various methods on aCGH data as simulated in \cite{louhimoetal}, without
subsampling.}\label{figsyntheticCNVfull}
\end{figure}

Figure \ref{figsyntheticCNVfull} reports comparisons of changepoint detection
and signal reconstruction accuracy for various methods on the original data
generated by the aCGH simulator of \cite{louhimoetal}; results for data
obtained by subsampling every 10th point of each sequence were reported in
Section \ref{subsecsyntheticCNV}.

\section{Preprocessing details for CNV analysis of the NCI-60 cell lines}
\label{sec:NCI60}
Our analyzed
data corresponds to measurements of the $\log_2$-intensity-ratio for the NCI-60
cell lines made using the Agilent human genome CGH oligonucleotide microarray
44B (GEO accession GPL11068), as reported in \citep{varmaetal} and publicly
available at \url{http://www.ncbi.nlm.nih.gov/geo/query/acc.cgi?acc=GSE48568}.
We discarded data for the PR:DU145(ATCC) and
PR:RC01 cell lines which were not part of the original NCI-60 DTP cell line
screen, yielding 125 sequences corresponding to 60 distinct cell lines. We
mapped microarray probe IDs to genomic locations using the annotation file
available at the Agilent website
\url{http://www.chem.agilent.com/cag/bsp/gene_lists.asp}.

As the samples do not correspond to the same gender, we discarded measurements
on the sex chromosomes. We observed a sizeable mean-shift of the entire data
sequence between replicate measurements of the same cell line, and hence
median-centered each sequence at 0.

The measurements of certain individual probes corresponded to large outliers in
the data sequences, with the outlier value being significantly higher
in some sequences and significantly lower in others. We believe such
measurements are likely due to technical noise in the Agilent oligonucleotide
platform, as previously noted in \cite{olshenetal} and \cite{nnowaketal}. We
applied an outlier removal procedure similar to that in \cite{olshenetal}:
For each sequence, we computed a median-absolute-deviation estimate of the noise
level $\sigma$. For each location $t$, if
the data value at $t$ was the maximum or minimum in the window
from $t-3$ to $t+3$, and if the difference between its
value and the closest other value in this window exceeded $2\sigma$, then
we replaced the value at $t$ with the median over this window.

\section*{Acknowledgements}
We would like to thank Ron Dror, David Siegmund, Janet Song, and Weijie Su for
helpful discussions and comments on an early draft of this paper. We would also
like to thank the referees and associate editor for suggestions that led to
many improvements in our data analyses.

\bibliographystyle{abbrvnat}
\bibliography{references}
\end{document}